\documentclass[aps,twocolumn,showpacs,floatfix]{revtex4-1}
\usepackage{amsfonts,amssymb,amsmath,subfigure,graphicx,color}
\begin{document}
\def\be{\begin{equation}}
\def\ee{\end{equation}}
\def\bea{\begin{eqnarray}}
\def\eea{\end{eqnarray}}
\def\fr{\frac}
\def\l{\label}
\def\om{\omega}
\def\eps{\epsilon}
\def\g{\gamma}
\def\bef{\begin{figure}[h!]}
\def\ef{\end{figure}}
\title{Linear response theory for long-range interacting systems in
quasistationary states}
\author{Aurelio Patelli,$^1$ Shamik Gupta,$^2$ Cesare Nardini,$^{1,2}$
Stefano Ruffo$^{2,3}$}
\affiliation{
$^1$ Dipartimento di Fisica ed Astronomia, Universit\`{a} di Firenze and INFN, Via Sansone 1, 50019 Sesto Fiorentino, Italy \\ 
$^2$ Laboratoire de Physique de l'\'{E}cole Normale Sup\'{e}rieure de Lyon,
Universit\'{e} de Lyon, CNRS, 46 All\'{e}e d'Italie, 69364 Lyon c\'edex 07, France \\
$^3$ Dipartimento di Energetica ``Sergio Stecco'' and CSDC,
Universit\`{a} di Firenze, CNISM and INFN, via S. Marta 3, 50139 Firenze, Italy}

\date{\today}
\begin{abstract}
Long-range interacting systems, while relaxing to equilibrium, often get
trapped in long-lived quasistationary states which have lifetimes
that diverge with the system size. 
In this work, we address the question of how a long-range system in a quasistationary
state (QSS) responds to an external perturbation. We consider a
long-range system that evolves under deterministic Hamiltonian dynamics.
The perturbation is 
taken to couple to the canonical
coordinates of the individual constituents. Our study is based on
analyzing 
the Vlasov equation for the single-particle phase-space distribution. The QSS
represents a stable stationary solution of the Vlasov equation in the absence
of the external perturbation. In the presence of small perturbation, we
linearize the perturbed Vlasov equation about the QSS to obtain a formal expression for the response observed in a
single-particle dynamical quantity. For a QSS that is homogeneous in the  
coordinate, we obtain an explicit formula for the response. We apply our
analysis to a
paradigmatic model, the Hamiltonian mean-field model, which
involves particles moving on a circle under Hamiltonian dynamics. Our
prediction for the response of three representative QSSs in this model (the water-bag
QSS, the Fermi-Dirac QSS, and the Gaussian QSS) is found to be in good agreement with $N$-particle simulations for
large $N$. We also show the long-time relaxation of the water-bag QSS to the Boltzmann-Gibbs
equilibrium state.
\end{abstract}
\maketitle
\section{Introduction}
\l{Introduction}
Systems with long-range interactions are ubiquitous in nature
\cite{review3,review4,review5}. In these
systems, the interaction potential between two particles decays
asymptotically with separation $r$ as $1/r^{\alpha}$, where $\alpha$ is smaller than or equal
to the spatial dimension of the system. Examples are dipolar
ferroelectrics and ferromagnets, self-gravitating
systems, non-neutral plasmas, two-dimensional geophysical
vortices, etc. Long-range interactions result in non-additivity, thereby
giving rise to equilibrium properties which are unusual for short-range systems, e.g., a negative microcanonical
specific heat, inequivalence of
statistical ensembles, and others \cite{review3}.

Long-range systems often exhibit intriguingly slow
relaxation toward equilibrium \cite{Yamaguchi:2004,Joyce:2010,Levin:2010,Gupta:2011}.
Such slow relaxation has been widely investigated in a model of globally
coupled particles moving on a unit circle and evolving under
Hamiltonian dynamics. This model is known as the Hamiltonian
mean-field (HMF) model \cite{Ruffo:1995}. In this model, a wide class of
initial conditions relaxes to equilibrium over times that diverge
with the
system size. It has been demonstrated that, depending on the initial
condition, the
relaxation to equilibrium for some energy interval occurs through intermediate long-lived
quasistationary states (QSSs). These
non-Boltzmann states involve slow evolution of thermodynamic
observables over time, and have a lifetime which grows algebraically with the
system size \cite{Yamaguchi:2004}. An immediate consequence is that the
system, in the limit of infinite size, never attains the Boltzmann-Gibbs equilibrium,
but remains trapped in the QSSs. 
Generalizations of the HMF model to include anisotropy terms in the energy \cite{Jain:2007}, and to particles
which are confined to move on a spherical surface rather than on a
circle \cite{Nobre:2003} have also shown slow relaxation
toward equilibrium and the presence of QSSs.

Dynamics of systems with long-range interaction, in the infinite-size
limit, is described by the Vlasov equation that governs the time evolution 
of the single-particle phase-space distribution \cite{Nicholson:1992}.
This equation allows a wide class of stationary solutions. Their
stability can be determined from the temporal behavior of small
fluctuations by linearizing the Vlasov equation around the stationary 
solutions. Stable stationary states of the Vlasov equation correspond to
QSSs of the finite-size system. The first clear demonstration of this
correspondence was achieved for the HMF model \cite{Yamaguchi:2004}.
In this paper, we refer interchangeably to stable
stationary states of the Vlasov equation and QSSs  
of the finite-size dynamics.

The study of the time evolution of fluctuations around a
stationary solution of the Vlasov dynamics is relevant in many
applications, for instance, in the phenomenon of
Landau damping that arises due to energy exchange between particles and waves in an electrostatic plasma \cite{Landau:1946}. 
In this process and in many others, {\em spontaneous} statistical
fluctuations
around a stable stationary state of the Vlasov equation
are considered, and their rate of exponential decay in time is determined
by solving an initial value problem involving the linearized Vlasov equation through a careful use of the Laplace-Fourier 
transform.

In this paper, we follow a different approach to the study of
fluctuations by the linearized 
Vlasov equation. Inspired by Kubo linear response theory
\cite{KuboJPSJ:1957}, we analyze the response of a Vlasov-stable stationary state to the application
of a small external perturbation described by a time-dependent
term in the Hamiltonian. The perturbation induces {\em forced} fluctuations around
the stationary state that we treat to linear order in the strength of
the perturbation, and study their evolution in time by using the
linearized Vlasov equation.
Such forced fluctuations are known to be generically finite for Boltzmann-Gibbs equilibrium states. 
We show here theoretically that they are finite and small, of the order of the perturbation,
also for Vlasov-stable stationary states. We support our analysis with $N$-particle numerical simulations
of the HMF model.

The paper is organized as follows. In Sec. \ref{LRT-general}, we develop
the linear response theory for a general QSS by using the Vlasov
framework.
In Sec. \ref{homQSS}, we specialize to a QSS that is homogeneous in the coordinate, and
derive a closed form expression for the change induced by the external perturbation in a
single-particle dynamical quantity. Section \ref{application-to-HMF} is devoted to the
application of the theory to study the response of three representative homogeneous QSSs in the HMF model, namely, the
widely studied water-bag QSS, the Fermi-Dirac QSS and the homogeneous
equilibrium state, which is also a QSS. In the following section, we
compare results from $N$-particle numerical simulations of the HMF dynamics
with those from the linear response theory, and obtain good
agreement. We also discuss the long-time relaxation of the
water-bag QSS to Boltzmann-Gibbs equilibrium under the action of the
perturbation.
We draw our conclusions in Sec. \ref{Conclusions}.

\section{Linear response theory for QSS}
\l{LRT-general}
Consider a system of $N$ particles interacting through a long-ranged
pair potential. The
Hamiltonian of the system is 
\be
H_0=\sum_{i=1}^N\fr{p_i^2}{2}+\fr{1}{N}\sum_{i<j}^N v(q_i-q_j),
\l{H0}
\ee
where $q_i$ and $p_i$ are, respectively, the canonical coordinate and
momentum of the $i$th particle, while $v(q_i-q_j)$ is the interaction
potential between the $i$th and $j$th particles. We assume that
$v(q)$ is suitably regularized for zero argument. We regard $q_i$'s and
$p_i$'s as one-dimensional variables, though our formalism may be
easily extended to higher dimensions. The mass of the particles is
taken to be unity. The factor $1/N$ in Eq.~(\ref{H0}) makes the energy extensive, in
accordance with the Kac
prescription \cite{Kac:1963}. In this work, we take unity for the
Boltzmann constant.
The system evolves under deterministic
Hamiltonian dynamics: the equations of motion for the $i$th particle are
\be
\dot{q}_i=p_i,~\dot{p}_i=-\fr{\partial}{\partial q_i}\fr{1}{N}\sum_{i<j}^N
v(q_i-q_j),
\l{EOM-unperturbed}
\ee
where dots denote differentiation with respect to time.

We start with the system in a quasistationary state (QSS) at time $t=0$, and
apply an external field $K(t)$. A QSS represents
a stable stationary
solution of the dynamics (\ref{EOM-unperturbed}) 
in the limit $N \rightarrow \infty$. For finite $N$, however, size
effects lead to instability and a slow relaxation of
the QSS to the Boltzmann-Gibbs
equilibrium state over a timescale that diverges with
$N$ \cite{Yamaguchi:2004,Joyce:2010,Levin:2010,Gupta:2011,Jain:2007,Nobre:2003}. 
 
Assuming the field $K(t)$ to couple to the coordinates of the individual particles, the perturbed Hamiltonian is 
\be
H(t)=H_0+H_{\rm ext}=H_0-K(t)\sum_{i=1}^Nb(q_i).
\l{H-perturbed}
\ee
Here, $b(q_i)$ denotes the dynamical quantity for the $i$th particle
that is conjugate to $K(t)$. The
equations of motion are modified from Eq.  
(\ref{EOM-unperturbed}) to
\be
\dot{q}_i=p_i,~\dot{p}_i=-\fr{\partial}{\partial
q_i}\fr{1}{N}\sum_{i<j}^N v(q_i-q_j)+K(t)\fr{\partial b(q_i)}{\partial
q_i}.
\l{EOM-perturbed} 
\ee

In this work, we study the temporal response of the initial QSS
to the field $K(t)$, in particular, the {\em linear} response. We
ask: How does a single-particle
dynamical quantity $a(q)$, that starts from a value corresponding to the
QSS, evolve in time under the action of $K(t)$? We seek
answers to this question by considering the system in the limit $N \to
\infty$, so that size effects are negligible and the evolution of the QSS is
 due to the field $K(t)$ alone. We also regard $K(t)$ to satisfy the
 following conditions: $K(t)$ is a monotonically increasing function of
 $t$ and has a value $\ll 1$ at all times, $K(t=0)=0$, and $K(t \to \infty)=$ a constant much
 smaller than $1$. While discussing the time-asymptotic response, we will mean the ordering of
limits $N \to \infty$ first, followed by $t \to \infty$.  
Note that the perturbed
dynamics Eq. (\ref{EOM-perturbed}) does not conserve the total energy of
the system as does Eq. (\ref{EOM-unperturbed}), although the variation
is expected to be small for small $K(t)$. 

The framework we adopt to address our queries is that of the Vlasov
equation for the time evolution of the single-particle phase space
distribution. For a system like Eq. (\ref{H0}) in the limit $N \to \infty$, such an equation faithfully describes the $N$-particle dynamics
in Eq. (\ref{EOM-unperturbed}) \cite{Braun:1977,Nicholson:1992}. That for small
$K(t)$ the Vlasov equation describes the perturbed dynamics of Eq. (\ref{EOM-perturbed}) in the infinite-size limit is illustrated later in
the paper by comparing the predictions of our analysis with $N$-particle
simulations for large $N$.

Let the function $f_d(q,p,t)$ count the
fraction of particles with coordinate $q$ and momentum $p$ at time $t$:
\be
f_d(q,p,t)=\frac{1}{N}\sum_{j=1}^N\delta
\Big(q-q_j(t)\Big)\delta\Big(p-p_j(t)\Big).
\l{empirical-measure}
\ee
In the limit $N \to \infty$, the function $f_d(q,p,t)$ converges to the smooth
function $f(q,p,t)$, which may be interpreted as the single-particle phase space distribution
function. The Vlasov equation
for the time evolution of $f(q,p,t)$ may be derived by using the
equations of motion (\ref{EOM-perturbed}) and following
standard approaches \cite{Nicholson:1992, Braun:1977}, and is given by
\be
\fr{\partial f}{\partial t}-\mathcal{L}(q,p,t)[f]f=0,
\l{Vlasov-equation}
\ee
where the operator $\mathcal{L}(q,p,t)[f]$ is given by  
\be
\mathcal{L}(q,p,t)[f]=-p\fr{\partial }{\partial q}+\fr{\partial \Phi(q,t)[f]}{\partial q}
\fr{\partial }{\partial p}-K(t)\fr{\partial b}{\partial q}\fr{\partial }{\partial p},
\l{Liouville-operator}
\ee
while $\Phi(q,t)[f]$ is the mean-field potential:
\be
\Phi(q,t)[f]=\iint dq'dp'~v(q-q')f(q',p',t).
\l{mean-field-potential}
\ee

We investigate the response of the system to the external field by
monitoring the observable
\be
\langle a(q) \rangle(t) \equiv \iint dq dp ~a(q)f(q,p,t).
\l{aqaverage-definition}
\ee
To obtain its time dependence, we need to solve Eq.~(\ref{Vlasov-equation}) for $f(q,p,t)$, with
the initial condition 
\be
f(q,p,0) = f_0(q,p).
\l{initial-condition}
\ee
Here, $f_0(q,p)$ characterizes a QSS, i.e., a stable stationary solution of the
Vlasov equation for the unperturbed dynamics (\ref{EOM-unperturbed}). Thus,
$f_0(q,p)$ satisfies
\be
\mathcal{L}_0(q,p)[f_0]f_0=0,
\l{QSS-definition}
\ee
where 
\be
\mathcal{L}_0(q,p)[f_0]=-p\fr{\partial }{\partial q}+\fr{\partial
\bar{\Phi}(q)[f_0]}{\partial q}\fr{\partial }{\partial p},
\l{L0}
\ee
and
\be
\bar{\Phi}(q)[f_0]=\iint dq'dp'~v(q-q')f_0(q',p').
\l{mean-field-potential-unperturbed}
\ee

To solve Eq.~(\ref{Vlasov-equation}) for $K(t) \ll 1$, we expand
$f(q,p,t)$ to linear order in $K(t)$ as
\be
f(q,p,t)=f_0(q,p)+\Delta f(q,p,t),
\l{f-decomposition}
\ee
with the initial condition
\be
\Delta f(q,p,0)=0.
\l{initial condition}
\ee

Substituting Eq.~(\ref{f-decomposition}) in Eq.~(\ref{Vlasov-equation}),
and separating terms to order $1$ and $K(t)$, we get, respectively, 
\bea
&&\fr{\partial f_0}{\partial t}-\mathcal{L}_0(q,p)[f_0]
f_0=0, \l{Vlasov-QSS}
\eea
and
\bea
&&\fr{\partial \Delta f}{\partial t}-\mathcal{L}_0(q,p)[f_0]\Delta
f=\mathcal{L}_{\rm ext}(q,p,t)[\Delta f]f_0. \l{Vlasov-linear} 
\eea 
Here, the operator
\be
\mathcal{L}_{\rm ext}(q,p,t)[\Delta f]=\fr{\partial \Phi(q,t)[\Delta f]}{\partial q}\fr{\partial }{\partial p}-K(t)\fr{\partial b}{\partial q}\fr{\partial }{\partial p}
\l{L-external}
\ee
describes the effects of the
external field, which are two-fold: (i) to generate a potential due to
its direct coupling with the
particles, and (ii) to modify the mean-field potential
(\ref{mean-field-potential}) from its value $\bar{\Phi}(q)[f_0]$ in the
absence of the field. Defining an effective single-particle potential,
\be
v_{\rm eff}(q,t)[\Delta f]=\Phi(q,t)[\Delta f]-K(t)b(q),
\l{veffective-definition}
\ee 
Eq.~(\ref{Vlasov-linear}) may be written as
\be
\fr{\partial \Delta f}{\partial t}-\mathcal{L}_0(q,p)[f_0]\Delta
f=\fr{\partial v_{\rm
eff}(q,t)[\Delta f]}{\partial q}\fr{\partial f_0}{\partial p}.
\l{Vlasov-linear-again}
\ee

Equation (\ref{Vlasov-QSS}) is satisfied by virtue of the
definition of $f_0(q,p)$. We thus solve Eq.~(\ref{Vlasov-linear-again})
for $\Delta f(q,p,t)$ in order to determine $f(q,p,t)$ from Eq.~(\ref{f-decomposition}).
With the condition (\ref{initial condition}), the formal solution is
\be
\Delta f(q,p,t)=\int\limits_0^t d\tau
~e^{(t-\tau)\mathcal{L}_0(q,p)[f_0]}\fr{\partial v_{\rm eff}(q,\tau)[\Delta
f]}{\partial q}\fr{\partial f_0(q,p)}{\partial p}.
\l{formal-solution-1}
\ee

Using Eq.~(\ref{formal-solution-1}) in
Eqs.~(\ref{aqaverage-definition}) and (\ref{f-decomposition})
gives the change in the value of $\langle
a(q) \rangle(t)$ due to the external field:
\bea
&&\langle \Delta a(q) \rangle(t)\equiv\iint dqdp~a(q)\Big(f(q,p,t)-f_0(q,p)\Big) \nonumber \\
&&=\int\limits_0^t d\tau \iint dqdp~a(q)
e^{(t-\tau)\mathcal{L}_0(q,p)[f_0]}\nonumber \\
&&\times \fr{\partial v_{\rm eff}(q,\tau)[\Delta
f]}{\partial q}\fr{\partial f_0(q,p)}{\partial p} \nonumber \\
&&=-\int\limits_0^t d\tau \Big\langle \fr{\partial
a(t-\tau)}{\partial p}\fr{\partial v_{\rm eff}(q,\tau)[\Delta
f]}{\partial q}\Big\rangle_{f_0}. 
\l{aq-average}
\eea
Here, angular brackets with $f_0$ in the subscript denote averaging with
respect to $f_0(q,p)$, e.g.,
\be
\langle a(q) \rangle_{f_0} \equiv \iint dq dp ~a(q)f_0(q,p),
\l{aqf0-average-definition}
\ee
while
\be
a(t-\tau)=e^{-(t-\tau)\mathcal{L}_0(q,p)[f_0]}a(q)
\ee
is the time-evolved $a(q)$ under the dynamics of the unperturbed system.
In obtaining the last equality in Eq.~(\ref{aq-average}), we have used the definition of
$\mathcal{L}_0$, 
have performed integration with respect to $q$, and have
assumed the boundary terms involving $f_0(q,p)$ to vanish. 

Defining the Poisson bracket between
two dynamical variables $g(q,p)$ and $g'(q,p)$ in the single-particle phase space as
\be
\{g(q,p),g'(q,p)\}\equiv\fr{\partial g}{\partial q}\fr{\partial
g'}{\partial p}-\fr{\partial g'}{\partial q}\fr{\partial g}{\partial p},
\l{poisson-bracket}
\ee
Eq.~(\ref{aq-average}) may be rewritten as
\bea
&&\langle \Delta a(q) \rangle(t) \nonumber \\
&&=\int\limits_0^t d\tau \Big\langle \{a(t-\tau),v_{\rm eff}(q,\tau)[\Delta
f]\}\Big\rangle_{f_0}. \l{Delta-a-again}
\eea
This is the central result of the paper. The above equation has a form similar to the Kubo
formula for the response of a dynamical quantity defined
in the full $2N$-dimensional phase space to an external perturbation
\cite{KuboJPSJ:1957}. The relation of formula (\ref{Delta-a-again}) with
more general ones derived by Ruelle \cite{Ruelle:2009} in the context of
dynamical system theory remains to be investigated.
In the following section, we discuss the special case of a homogeneous QSS, i.e., $f_0(q,p)=P(p)$ is a function solely of the
momentum, to obtain an explicit
form of the formal solution (\ref{formal-solution-1}).
\section{Homogeneous QSS}
\l{homQSS}
We consider a homogeneous QSS with
$f_0(q,p)=P(p)$, where $P(p)$ is any distribution
of the momentum, with the normalization
\be
\int dqdp~ P(p)=1, ~~~~\int dp~ P(p)=\fr{1}{V},
\ee
where
\be
V \equiv \int dq
\ee
is the total volume of the coordinate space.

For a homogeneous QSS, Eq.~(\ref{L0}) gives
\be
\mathcal{L}_0(q,p)[f_0]=-p\fr{\partial }{\partial q},
\l{L0-homQSS}
\ee
so that Eq.~(\ref{formal-solution-1}) becomes
\be
\Delta f(q,p,t)=\int\limits_0^t d\tau
~e^{-(t-\tau)p\fr{\partial }{\partial q}}\fr{\partial v_{\rm eff}(q,\tau)[\Delta
f]}{\partial q}\fr{\partial P(p)}{\partial p},
\l{formal-solution-homQSS}
\ee
which implies that the spatial Fourier
and temporal Laplace transform of $\Delta f(q,p,t)$ satisfies
\cite{FourierLaplace} 
\bea
&&\widehat{\Delta f}(k,p,\om)=\fr{\partial P(p)}{\partial
p}ikL[e^{-tpik}]\nonumber \\
&&\times \Big[2\pi\widetilde{v}(k)\int dp' 
~\widehat{\Delta f}(k,p',\om) 
-\widehat{K}(\om)\widetilde{b}(k)\Big]. 
\l{Fourier-Laplace-solution}
\eea
Here, $L$ denotes the Laplace transform: 
\be
L[e^{-tpik}]=\int_0^\infty dt~e^{i\om t-tpik}=\fr{1}{i(kp-\om)},
\ee
assuming Im$(\om)$ to be positive.
Thus, 
\bea
&&\widehat{\Delta f}(k,p,\om)\nonumber \\
&&=
\fr{\partial P(p)}{\partial
p}\fr{k}{kp-\om}\Big[2\pi\widetilde{v}(k)\int dp' 
~\widehat{\Delta f}(k,p',\om) 
-\widehat{K}(\om)\widetilde{b}(k)\Big]. \nonumber \\
\l{Fourier-Laplace-solution-1}
\eea

Now, integrating both sides of Eq.~(\ref{Fourier-Laplace-solution-1}) with
respect to $p$ gives
\be
\int dp~\widehat{\Delta f}(k,p,\om) 
=\fr{\widehat{K}(\om)\widetilde{b}(k)}{2\pi
\widetilde{v}(k)}\Big(\fr{\eps(k,\om)-1}{\eps(k,\om)}\Big).
\l{final-1}
\ee
Here, $\eps(k,\om)$ is the ``dielectric function"
\cite{review3}:
\be
\eps(k,\om)=1-2\pi k\widetilde{v}(k)\int_{LC} \fr{dp}{kp-\om} \fr{\partial
P(p)}{\partial p},
\l{dielectric-function-definition}
\ee
where the integral has to be performed along the Landau contour $LC$
that makes Eq. (\ref{final-1}) valid in the whole of
the $\om$-plane; we have \cite{Nicholson:1992,Chavanis:2009}:
\bea
\eps(k,\om)=\left\{ 
\begin{array}{l}
                1-2\pi k \widetilde{v}(k)\int\limits_{-\infty}^\infty
                \fr{dp}{kp-\om} \fr{\partial
P(p)}{\partial p} \\
~~~~~~~~~~~~~~~~~~~~({\rm Im}(\om) >0), \\ \\
               1-2\pi k \widetilde{v}(k){\rm
               P}\int\limits_{-\infty}^\infty \fr{dp}{kp-\om} \fr{\partial P(p)}{\partial
p}\\-i2\pi^2  \widetilde{v}(k) \left.\fr{\partial P(p)}{\partial
p}\right|_{\om/k} \\
~~~~~~~~~~~~~~~~~~~~({\rm Im}(\om) =0), \nonumber \\ \\
               1-2\pi k \widetilde{v}(k)\int\limits_{-\infty}^\infty
               \fr{dp}{kp-\om} \fr{\partial P(p)}{\partial
p}\\-i4\pi^2  \widetilde{v}(k) \left.\fr{\partial P(p)}{\partial
p}\right|_{\om/k}\\
~~~~~~~~~~~~~~~~~~~~({\rm Im}(\om) < 0), \nonumber \\
               \end{array}
        \right. \\
\l{eps-explicit}
\eea
where ${\rm P}$ denotes the principal part.

From Eq.~(\ref{f-decomposition}), the change in the
distribution due to the external field is 
\bea
f(q,p,t)-P(p)=\fr{1}{2\pi}\int_C d\om ~e^{-i\om t}\int
dk~e^{ikq}\widehat{\Delta f}(k,p,\om), \nonumber \\ 
\eea
where $C$ is the Laplace contour.
Integration over $p$ gives
\bea
&&\int dp ~f(q,p,t)=\fr{1}{V} \nonumber \\
&&+\fr{1}{2\pi}\int_C d\om ~e^{-i\om t}\int dk~e^{ikq}\Big[\fr{\widehat{K}(\om)\widetilde{b}(k)}{2\pi
\widetilde{v}(k)}\Big(\fr{\eps(k,\om)-1}{\eps(k,\om)}\Big)\Big],
\nonumber \\ \l{final-complete-solution-1}
\eea
where we have used Eq.~(\ref{final-1}).
Let us suppose that the expression enclosed
by square brackets has singularities which are isolated poles of any
order. Let $\{\om_{\rm p}(k)\}$ be the set of poles, while $\{r_{\rm
p}(k)\}$ is the set of residues at these poles. Then, by the theorem of residues, we get
\be
\int dp~f(q,p,t)=\fr{1}{V}+\fr{1}{2\pi}\int dk~e^{ikq}\sum_{\rm p}  (2\pi
i)r_{\rm p}(k) e^{-i\om_{\rm p}(k) t}.
\l{residue}
\ee

From Eq.~(\ref{final-complete-solution-1}), we see that the poles
correspond either to poles of $\widehat{K}(\om)$ or to the zeros of the dielectric function, i.e., values 
$\om_{\rm p}(k)$ (complex in general) that satisfy
\be
\eps(k,\om_{\rm p}(k))=0.
\l{dispersion-relation-1}
\ee
Equation (\ref{residue}) implies that these poles determine the growth or decay of the difference $\int dp~f(q,p,t)-\fr{1}{V}$ in time depending on
the location of the poles in the complex-$\om$ plane. For example, when there are
poles in the upper-half complex $\om$-plane, the difference grows in time. If,
on the other hand, the poles lie either on or below the real-$\om$ axis,
the difference does not grow in time, but oscillates or decays
in time, respectively. 

We have to ensure that our analysis leading to Eq.~(\ref{residue}) is
consistent with the decomposition in 
Eq. (\ref{f-decomposition}) for perturbations about a
{\em stable} stationary state $f_0(q,p)=P(p)$. It is thus required that 
$\int dp~f(q,p,t)-\fr{1}{V}$ does not
grow in time, which means that the aforementioned poles cannot lie in
the upper-half $\om$-plane. Now, since $K(t)$ was chosen to satisfy the
conditions $K(t=0)=0$ and $K(t \to \infty)=$ a constant much
 smaller than $1$, it follows that $\widehat{K}(\om)$ cannot have poles in the upper-half $\om$-plane. We therefore conclude that Eq.~(\ref{residue}) is valid
when the poles $\om_{\rm p}(k)$ that come from the zeros of $\eps(k,\om)$
satisfy 
\be
\eps(k,\om_{\rm p}(k))=0; ~~~~{\rm Im}(\om_{\rm p}(k)) \le 0,
\l{dispersion-relation}
\ee
corresponding to linear stability of the stationary state $P(p)$.
The condition ${\rm Im}(\om_{\rm p}(k)) = 0$ corresponds to marginal
stability of $P(p)$. In this case, the zeros of the dielectric function lie on the
real-$\om$ axis so that $\om_{\rm p}(k)=\om_{\rm pr}(k)$ is real. From
Eqs.~(\ref{dispersion-relation-1}) and  
(\ref{eps-explicit}), we find
that $\om_{\rm pr}(k)$ satisfies 
\bea
&&1-2\pi k \widetilde{v}(k){\rm
P}\int_{-\infty}^\infty \fr{dp}{kp-\om_{\rm pr}(k)} \fr{\partial P(p)}{\partial
p}\nonumber \\
&&-i2\pi^2  \widetilde{v}(k)\left.\fr{\partial P(p)}{\partial
p}\right|_ {\om_{\rm pr}(k)/k}=0.
\l{dielectric-marginal}
\eea
Equating the real and the imaginary parts to zero,
we get
\bea
&&1-2\pi k \widetilde{v}(k) {\rm P}\int_{-\infty}^\infty \fr{dp}{kp
-\om_{\rm pr}(k)} \fr{\partial P(p)}{\partial
p}=0, \l{marginal-stability-real-part}\\
&&\left.\fr{\partial P(p)}{\partial
p}\right|_ {\om_{\rm pr}(k)/k}=0.\l{marginal-stability-im-part}
\eea

We now move on to apply our analysis to the Hamiltonian mean-field
model, a paradigmatic model of long-range interactions.
\section{Application to the Hamiltonian mean-field model}
\l{application-to-HMF}
\subsection{Model}
\l{HMF-model}
The Hamiltonian mean-field (HMF) model belongs to the
class of models with Hamiltonian
(\ref{H0}), with the additional feature that the potential
$v(q)$ is even:
\be
v(q)=1-\cos q, 
\l{HMF-vq}
\ee
so that the Hamiltonian is \cite{Ruffo:1995}
\be
H_0=\sum_{i=1}^N\fr{p_i^2}{2}+\fr{1}{N}\sum_{i<j}^N \left[ 1-\cos
(q_i-q_j) \right].
\l{HMF-Hamiltonian}
\ee
The model describes particles moving on a unit circle under 
Hamiltonian dynamics (\ref{EOM-unperturbed}). The
canonical coordinate $q_i \in [0,2\pi]$ specifies the angle for the location of the
$i$th particle on the circle with respect to an arbitrary fixed axis, while
$p_i$ is the conjugate momentum \cite{noteXYspins}.

The model in the equilibrium state shows a continuous transition from a
low-energy clustered phase, in which the particles are close together on
the circle, to a high-energy
homogeneous phase corresponding to a uniform distribution of
particles on the circle. The clustering of the particles is measured by the magnetization vector $\langle \mathbf{m}\rangle(t)$ with components
\be
(\langle m_x \rangle(t), \langle m_y \rangle(t))=\iint dqdp ~(\cos q,
\sin q)f(q,p,t),
\l{mx-my}
\ee
and magnitude $\langle m \rangle(t)=\sqrt{\langle m_x\rangle^2(t)+\langle m_y
\rangle^2(t)}$. In terms of $\langle m \rangle(t)$,
the energy density is 
\be
e=\Big\langle\fr{p^2}{2}\Big\rangle(t)+\fr{1}{2}\Big[1-\langle m
\rangle^2(t)\Big],
\l{HMF-e-definition}
\ee
where the kinetic energy $\Big\langle\fr{p^2}{2}\Big\rangle(t)$ defines the temperature $T$ of the system:
\be
\Big\langle\fr{p^2}{2}\Big\rangle(t)=\iint dqdp
~\fr{p^2}{2}f(q,p,t)=\fr{T}{2}.
\ee
Note that $e$ is conserved under the dynamics.

In equilibrium, the single-particle distribution
assumes the canonical form, $f_{\rm eq}(q,p)$, which is Gaussian in $p$
with a non-uniform distribution for $q$ below the transition energy density
$e_c$ and a uniform one above 
\cite{Rapisarda:1999}:
\be
f_{\rm eq}(q,p)=\frac{\sqrt{\beta}\exp\Big[-\beta \left( \frac{p^2}{2} 
- m_x^{eq}\cos (q -\phi) \right)\Big]}{(2 \pi)^{3/2} I_0(\beta m_x^{eq})}.
\l{HMF-Gaussian-solution}
\ee 
Here, $I_0$ is the modified Bessel function of zero order, $\beta$ is
the inverse temperature, while $m_x^{eq}$ is the equilibrium magnetization that decreases continuously from unity at zero 
energy density to zero at $e_c$ and remains zero at higher energies. The
arbitrary phase $\phi$ in Eq.~(\ref{HMF-Gaussian-solution}) is a
result of the rotational invariance of the Hamiltonian
(\ref{HMF-Hamiltonian}).
The energy at equilibrium is 
\be
e=\frac{1}{2 \beta}+\frac{1-(m_x^{eq})^2}{2}.
\ee

The phase transition in the HMF model 
occurs within both microcanonical
and canonical ensembles \cite{Ruffo:1995,Barre:2005}. Thus, ensemble equivalence, though not
guaranteed for long-range interacting systems, holds for the HMF model
\cite{review3}. The microcanonical transition energy is $e_c=3/4$, which corresponds 
to a transition temperature $T_c=1/2$ in the canonical ensemble. 

\subsection{Linear response of homogeneous QSS}
\l{linear-response-HMF}
Consider the QSS distribution $f_0(q,p)=P(p)$ which is
homogeneous in coordinate (thus, $\langle m_x
\rangle_{f_0}=\langle m_y \rangle_{f_0}=0$), but has an arbitrary normalized
distribution for the momentum. Here, we study the
response of this QSS to the external perturbation
\be
H_{\rm ext}=-K(t)\sum_{i=1}^N \cos q_i,
\l{perturbation-HMF}
\ee
which corresponds to the choice 
\be
b(q)=\cos q.
\l{bq-HMF}
\ee
in Eq.~(\ref{H-perturbed}).
The specific $K(t)$ we choose is a step function:
\be
K(t) =
\left\{\begin{array}{l l l}
    0 &{\rm for~} t<0, \\
     h  & {\rm for~} t\geq 0; ~~~~ h \ll 1. 
    \end{array}\right.
  \l{HMF-perturbation-Kt}
\ee

The changes in the magnetization components due to the field are 
\bea
&&\hspace{-0.4cm}\langle \Delta m_x \rangle(t)=\iint dqdp ~\Big(f(q,p,t)-P(p)\Big)\cos q 
\nonumber \\
&&\hspace{-0.4cm}=\fr{1}{2}\int_C d\om ~e^{-i\om t}\int dp ~\Big(\widehat{\Delta
f}(1,p,\om)+\widehat{\Delta f}(-1,p,\om)\Big), 
\l{mx}
\eea
and
\bea
&&\hspace{-0.4cm}\langle \Delta m_y \rangle(t)=\iint dqdp~ \Big(f(q,p,t)-P(p)\Big)\sin q
\nonumber \\
&&\hspace{-0.4cm}=\fr{1}{2i}\int_C d\om~e^{-i\om t}\int dp ~\Big(\widehat{\Delta
f}(-1,p,\om)-\widehat{\Delta f}(1,p,\om)\Big). \l{my}
\eea

Using 
\be
\widetilde{v}(k)=\Big[\delta_{k,0}-\fr{\delta_{k,-1}+\delta_{k,1}}{2}\Big],
\l{vk-HMF}
\ee
\be
\widetilde{b}(k)=\fr{\delta_{k,-1}+\delta_{k,1}}{2},
\l{bk-HMF}
\ee
\be
\widehat{K}(\om)=-\fr{h}{i\om},
\l{HMF-Komega}
\ee
and Eq. (\ref{final-1}) in Eqs. (\ref{mx}) and (\ref{my}) gives
\bea
\langle\Delta  m_x \rangle(t)=\fr{h}{2\pi}\int_C d\om~e^{-i\om
t}\fr{1}{i\om}\Big(\fr{\eps(1,\om)-1}{\eps(1,\om)}\Big),
\l{mx-1-h1}
\eea
and
\be
\langle \Delta m_y \rangle(t)=0.
\l{my-1-h1}
\ee
Here, we have used the fact that for the HMF model,
\be
\eps(1,\om)=\eps(-1,\om),
\ee
as may be easily checked by using Eq. (\ref{vk-HMF}) in Eq. (\ref{dielectric-function-definition}).
It may also be seen that
\be
\eps(k,\om)=1 {\rm ~for~} k \ne \pm 1.
\l{eps-HMF-kne1}
\ee

Now, using the fact that $\langle m_x
\rangle_{f_0}=\langle m_y \rangle_{f_0}=0$, Eqs.~(\ref{mx-1-h1}) and
(\ref{my-1-h1}) imply that 
\bea
\langle  m_x \rangle(t)=\fr{h}{2\pi}\int_C d\om~e^{-i\om
t}\fr{1}{i\om}\Big(\fr{\eps(1,\om)-1}{\eps(1,\om)}\Big),
\l{mx-1-h}
\eea
and
\be
\langle m_y \rangle(t)=0.
\l{my-1-h}
\ee
It can be proven straightforwardly from the Vlasov equation
(\ref{Vlasov-equation}) that Eq. (\ref{my-1-h}) holds also in the non-linear
response regime ($K(t)$ not necessarily small) for all homogeneous
$P(p)$ which are even in $p$. 

When the zeros of $\eps(1,\om)$ lie only in the lower-half complex-$\om$
plane, Eq.~(\ref{mx-1-h}) gives the time-asymptotic response: 
\bea
\overline{m}_x\equiv\lim_{t\to \infty}\langle m_x
\rangle(t)=h\Big(\fr{1-\eps(1,0)}{\eps(
1,0)}\Big). \l{mx-h-2} 
\eea
This equation implies diverging magnetization at $\eps(1,0)=0$, which is
clearly not possible as the magnetization is bounded above by unity.
Therefore, in such a case, the linear response theory makes an incorrect
prediction. Thus, we rely on
formula (\ref{mx-h-2}) only when the result is much smaller than unity.

Note that $\widehat{K}(\om)$, given in Eq.~(\ref{HMF-Komega}), has a
pole only at $\om=0$. Following the discussions in Sec.~\ref{homQSS}, we
thus conclude that the conditions (\ref{marginal-stability-real-part}) and
(\ref{marginal-stability-im-part}) solely determine the parameters
characterizing the distribution $P(p)$ such that it is marginally stable. For the HMF model, we need to consider only $k=\pm 1$ in these
conditions. Since $\eps(1,\om)=\eps(-1,\om)$, we
write $\om_{\rm pr}(1)=\om_{\rm pr}(-1)=\om_{\rm pr}$, so that these
conditions become
\bea
&&1+\pi{\rm P}\int_{-\infty}^\infty \fr{dp}{p\mp \om_{\rm pr}} \fr{\partial P(p)}{\partial
p}=0, \l{marginal-stability-real-part-HMF}\\
&&\left.\fr{\partial P(p)}{\partial
p}\right|_{\om_{\rm pr}}=0.\l{marginal-stability-im-part-HMF}
\eea   

We now consider two representative $P(p)$ and obtain the
linear response of the corresponding QSS by using Eq.~(\ref{mx-1-h}).
For the first case, we obtain the full temporal
behavior of the response, while in the second case, we discuss only the time-asymptotic response. 
\subsubsection{Water-bag QSS}
\l{waterbag-QSS}
The water-bag state
corresponds to coordinates uniformly distributed in $[0,2\pi]$ and
momenta uniformly
distributed in $[-p_0,p_0]$:
\be
P(p)=\fr{1}{2\pi}\fr{1}{2p_0}\Big[\Theta(p+p_0)-\Theta(p-p_0)\Big]; ~~~p \in
[-p_0,p_0].
\l{distribution-WB}
\ee
Here, $\Theta(x)$ denotes the unit step function.
The energy density is obtained from Eq.~
(\ref{HMF-e-definition}) as
\be
e=\fr{p_0^2}{6}+\fr{1}{2}.
\l{e-WB}
\ee
The dielectric function may be obtained
straightforwardly by using Eq.~(\ref{dielectric-function-definition}) to get
\be
\eps(1,\om)=1-\fr{1}{2(p_0^2-\om^2)},
\l{eps-WB-omega}
\ee
which is analytic in the whole of the $\om$-plane, except at the two
points $\om=\pm p_0$.

As discussed in Sec. \ref{homQSS}, the zeros of the dielectric function determine the temporal
behavior of the difference $\int dp~f(q,p,t)-1/V$ (here $V=2\pi$). The zeros of
Eq. (\ref{eps-WB-omega}) occur at $\om_{\rm p}=\pm \sqrt{p_0^2-1/2}$. For
$p_0 < p_0^*=1/\sqrt{2}$, (correspondingly, $e < e^*=7/12$), the pair of zeros lies on the imaginary-$\om$ axis,
one in the upper half-plane and one in the lower. The one in the upper
half-plane makes the water-bag state linearly unstable for $e
< e^*$. As $e$ approaches $e^*$ from below, the zeros move along the
imaginary-$\om$ axis and hit the origin when $e=e^*$. At higher energies,
the zeros start moving on the real-$\om$ axis away from the origin in
opposite directions. The fact that the zeros of the dielectric function
are strictly real for $e \ge e^*$ implies that the water-bag state is
marginally stable at these energies, and is therefore a QSS at these
energies. 

From the discussions in Sec.~\ref{homQSS} and those following Eq.
(\ref{mx-h-2}), it follows that the result of the linear Vlasov
theory, Eq. (\ref{mx-1-h}), is valid and physically meaningful only when $p_0^2>1/2$.
Using Eq.~
(\ref{eps-WB-omega}) in  Eq.~(\ref{mx-1-h}) and performing the integral
by the residue theorem gives
\be
\langle m_x
\rangle(t)=\fr{2h}{2p_0^2-1}\sin^2\Big(\fr{t}{2}\sqrt{p_0^2-\fr{1}{2}}\Big);
~~~~p_0^2 > \fr{1}{2}.
\l{mx_WB_vs_time}
\ee
Thus, the linear Vlasov theory predicts that in the presence of an
external field along $x$, the corresponding magnetization exhibits oscillations 
for all times and does not approach any time-asymptotic constant value. This 
prediction is verified in numerical simulations discussed in 
Sec.~\ref{Single-initial-state}.
The average of $\langle m_x
\rangle(t)$ over a period of oscillation is
\be
\langle m_x \rangle_{\rm Time~ average} \equiv \fr{1}{T} \int_0^T dt ~\langle m_x \rangle (t)=\fr{h}{2 p_0^2-1};~~~~p_0^2 > \fr{1}{2},
\l{timeaverage}
\ee
where $T$ is the period of oscillation. In Sec.~\ref{simulations},
we will compare this average with numerical results.

\subsubsection{Fermi-Dirac QSS}
\l{Fermi-Dirac-QSS}
We now consider a Fermi-Dirac state in which the
coordinate is uniformly distributed in $[0,2\pi]$, while the
momentum has the usual Fermi-Dirac distribution:
\be
P(p)=A\fr{1}{2\pi}\fr{1}{1+e^{\beta(p^2-\mu)}}; ~~~p \in
[-\infty,\infty].
\l{distribution-FD}
\ee
Here, $\beta \ge 0$ and $\mu \ge 0$ are parameters characterizing the distribution,
while $A$ is the normalization constant. We consider the state
(\ref{distribution-FD}) in the limit of large $\beta$ in which analytic
computations of various physical quantities is possible. As $\beta \to \infty$ the
Fermi-Dirac state converges to the water-bag state (\ref{distribution-WB})
with $p_0=\sqrt{\mu}$.

As shown in the Appendix, to leading order in $1/\beta^2$, the
normalization is given by 
\be
A=\fr{1}{2\sqrt{\mu}}\Big(1+\fr{\pi^2}{24\beta^2\mu^2}\Big),
\l{norm-FD-final-text}
\ee
while the energy density is
\be
e=\fr{\mu}{6}\Big(1+\fr{\pi^2}{6\beta^2\mu^2}\Big)+\fr{1}{2}.
\l{energy-FD-final-text}
\ee

Let us now investigate the conditions
(\ref{marginal-stability-real-part-HMF}) and (\ref{marginal-stability-im-part-HMF}) for the marginal
 stability of the state (\ref{distribution-FD}).
Since $P(p)$ satisfies $\left.\fr{\partial P(p)}{\partial
p}\right|_{p=0}=0$, the
 condition (\ref{marginal-stability-im-part-HMF}) implies that
$\om_{\rm pr}=0$, which on substituting in condition
(\ref{marginal-stability-real-part-HMF}) gives
\be
\eps(1,0)=0,
\l{stability-FD-again}
\ee
where, as shown in the Appendix, to order $1/\beta^2$, we have
\be
\eps(1,0)=1-\fr{1}{2\mu}\Big(1+\fr{\pi^2}{6\beta^2\mu^2}\Big).
\l{mustar-FD}
\ee
Solving Eq.~
(\ref{stability-FD-again}) gives $\mu^*$, the value of $\mu$ at the marginal stability of
the state (\ref{distribution-FD}). To order $1/\beta^2$, we get
\be
\mu^*=\fr{1}{2}+\fr{2\pi^2}{3\beta^2},
\l{mustar-FD-final}
\ee
which gives the corresponding energy density 
\be
e^*=\fr{7}{12}+\fr{\pi^2}{6\beta^2},
\l{estar-FD-final}
\ee
such that at higher energies, the state (\ref{distribution-FD}) is 
a QSS.

Following our earlier discussions on the regime of validity of the linear Vlasov
theory, and using Eq.~(\ref{mustar-FD}) in Eq.~(\ref{mx-h-2}), we get
\be
\overline{m}_x=\fr{h\Big(1+\fr{\pi^2}{6\beta^2\mu^2}\Big)}{2\mu-1-\fr{\pi^2}{6\beta^2\mu^2}};
~~~~\mu > \mu^*.
\l{mx-final-FD}
\ee
\subsection{Linear response of the homogeneous equilibrium state}
\l{HMF-lrt}
It is interesting to consider the response of the distribution
(\ref{HMF-Gaussian-solution}) with magnetization $m_x^{eq}=0$,
which is the equilibrium state of the HMF model for energies $e > e_c$.
We thus consider the choice
\be
P(p)=\sqrt{\fr{\beta}{2\pi}}\exp\Big[-\frac{\beta p^2}{2}\Big].
\l{distribution-Gaussian}
\ee 

It is known that the equilibrium state (\ref{distribution-Gaussian}) is also a QSS
\cite{review3}. Indeed, stability condition (\ref{marginal-stability-im-part-HMF}) gives
$\om_{\rm pr}=0$, so that Eq.~(\ref{marginal-stability-real-part-HMF})
gives
\be
\eps(1,0)=0,
\ee
where $\eps(1,0)$ is given by
\be
\eps(1,0)=1-\fr{\beta}{2}.
\l{eps-equilibrium}
\ee
Thus, the state
(\ref{HMF-Gaussian-solution}) is marginally stable at $\beta=2$,
and correspondingly, $e=e^*=3/4=e_c$. For $e >
e^*$, the state is a QSS and also the Boltzmann-Gibbs
equilibrium state.

Using Eqs.~(\ref{eps-equilibrium}) and (\ref{mx-h-2}), one gets
\be
\overline{m}_x=\fr{h}{2/\beta -1};~~~~\beta < 2.
\l{mx-final-equilibrium}
\ee

Therefore, under the perturbation, Eqs.~(\ref{perturbation-HMF}) and (\ref{HMF-perturbation-Kt}), the
equilibrium state evolves to an inhomogeneous QSS predicted by our
linear response theory. Let us compare the value of $\overline{m}_x$ in Eq.~(\ref{mx-final-equilibrium})
with the one predicted by equilibrium statistical mechanics,
$m_x^{eq}(\beta,h)$, at the same values of the energy and $h$. This latter quantity is
obtained by solving the implicit equation
\cite{review3},
\be
\fr{X}{\beta}-h=\fr{I_1(X)}{I_0(X)},
\l{consistency}
\ee
with $I_1(X)$ the modified Bessel function of first order,
and using the solution $\bar{X}(\beta,h)$ to get
\be
m_x^{eq} (\beta,h)= \frac{I_1(\bar{X})}{I_0(\bar{X})}.
\l{eqmx}
\ee
The corresponding energy is
\be
e=\frac{1}{2 \beta}+\frac{1-(m_x^{eq}(\beta,h))^2}{2}-h m_x^{eq}(\beta,h).
\ee

The two values given in Eqs.~(\ref{mx-final-equilibrium}) and (\ref{eqmx})
are in general different. However, in the high-energy regime, one can solve Eq. (\ref{consistency}) for small
$X$ to obtain for the equilibrium magnetization the same formula as the
one obtained by the linear response theory, Eq. (\ref{mx-final-equilibrium}).
While comparing the two magnetization values with numerical results at
high energies in
Sec.~\ref{Linear-response-equilibrium-numerical}, we are thus not able
to distinguish between equilibrium and QSS magnetization in the presence
of the field.
\section{Comparison with $N$-particle simulations}
\l{simulations}

To verify the analysis presented in Sec.~\ref{application-to-HMF}, we
performed extensive numerical simulations of the $N$-particle dynamics (\ref{EOM-perturbed}) 
for the HMF model for large $N$. The equations of motion were integrated 
using a fourth-order symplectic scheme \cite{McLachlan:1992}, with a time step 
varying from $0.01$ to $0.1$.
In simulations, we prepare the HMF system at time $t=0$ in an initial state by sampling independently for every
particle the coordinate $q$ uniformly in $[0,2\pi]$ and the momentum
according to either the water-bag, the Fermi-Dirac, or the Gaussian
distribution. Thus, the probability distribution of the initial state is
\be
P(q_1,p_1,q_2,p_2,\ldots,q_N,p_N)=\prod_{i=1}^N P(p_i)
\l{initial-state}
\ee
where $P(p)$ is given by either Eq. (\ref{distribution-WB}),
(\ref{distribution-FD}), or (\ref{distribution-Gaussian}). The energy of
the initial state is chosen to be such that it is a QSS. Then, at time
$t_0 > 0$, we switch on the external perturbation, Eqs. (\ref{perturbation-HMF}) and
(\ref{HMF-perturbation-Kt}), and follow
the time evolution of the $x$-magnetization.

In obtaining numerical results, two different approaches were
adopted. In one approach, we followed in time the evolution of a {\it single realization} of the initial
state. These simulations are intended to check if our predictions based
on the Vlasov equation for the smooth distribution $f(q,p,t)$ for
infinite $N$ are also valid for a {\it typical} 
time-evolution trajectory of the empirical measure $f_d(q,p,t)$ for finite $N$, where
the initial condition $f_d(q,p,0)$ is obtained from Eqs.
(\ref{initial-state}) and (\ref{empirical-measure}), while $f(q,p,0)=P(p)$.
Rigorous results from Braun and Hepp and further analysis by Jain {\em
et al.} show that these typical trajectories stay close 
to the trajectory of $f(q,p,t)$ for times that increase logarithmically with $N$
\cite{Braun:1977,Jain:2007}. When $P(p)$ is a stable stationary solution
of the Vlasov equation, it is known numerically \cite{Yamaguchi:2004} and analytically
\cite{Caglioti:2008} that these times diverge as a power of $N$, and are
therefore sufficiently long to allow us to check even for moderate values of $N$ the predictions of our
linear Vlasov theory for perturbations about Vlasov-stable
stationary solution $P(p)$. 

In a second approach, we
obtained numerical results by averaging over an {\it ensemble of
realizations} of the initial state. The time evolution that we get using this second method 
is different from the first one.  This approach allows us to reach
the average and/or asymptotic value of an observable, here $\langle m_x
\rangle(t)$, on a faster
time scale because of a mechanism of convergence in time, as we
describe below.

\subsection{Linear response of homogeneous QSS: Single realization}
\l{Single-initial-state}

\begin{figure}[h!]
\centering
\includegraphics[width=90mm]{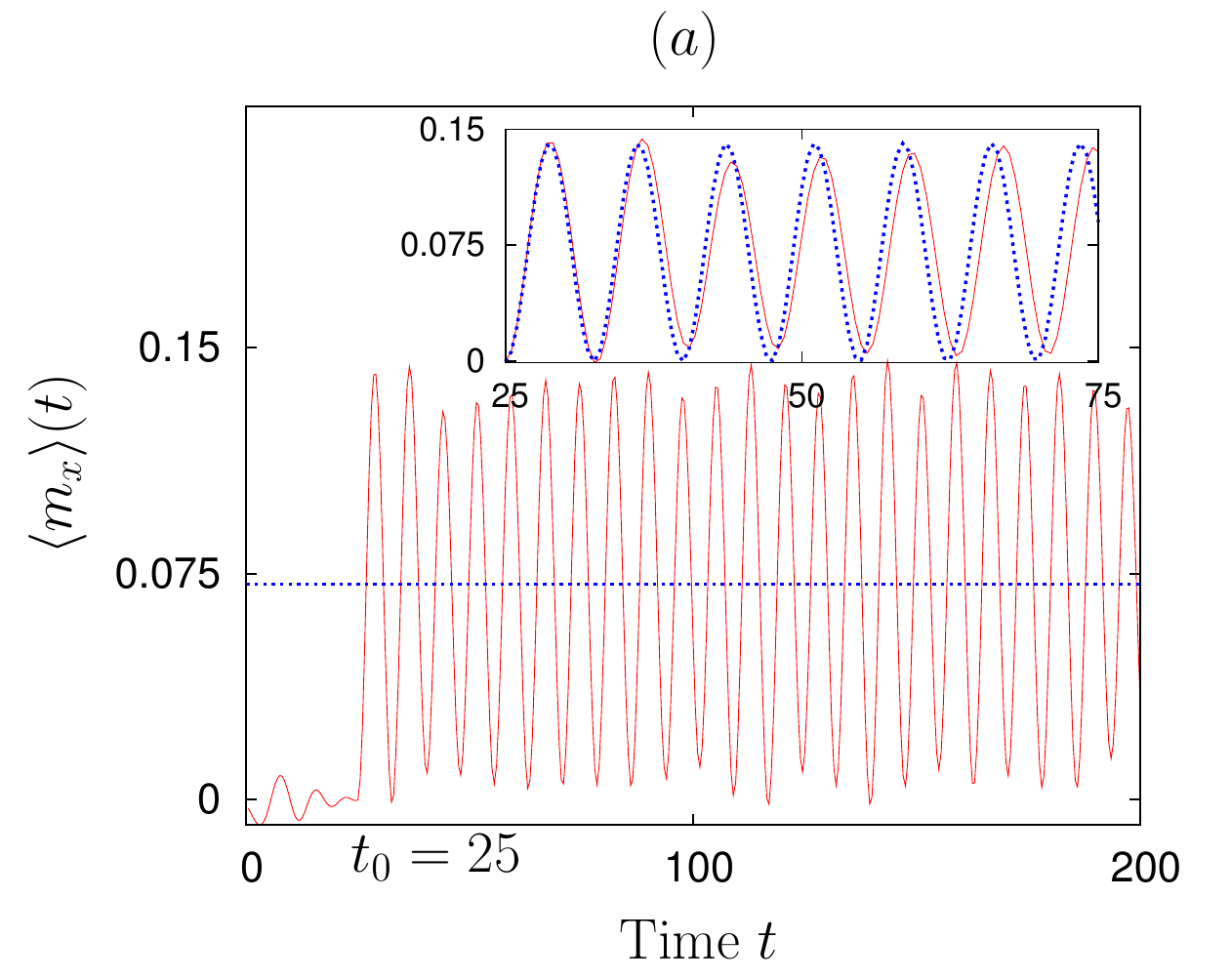}\\
\includegraphics[width=90mm]{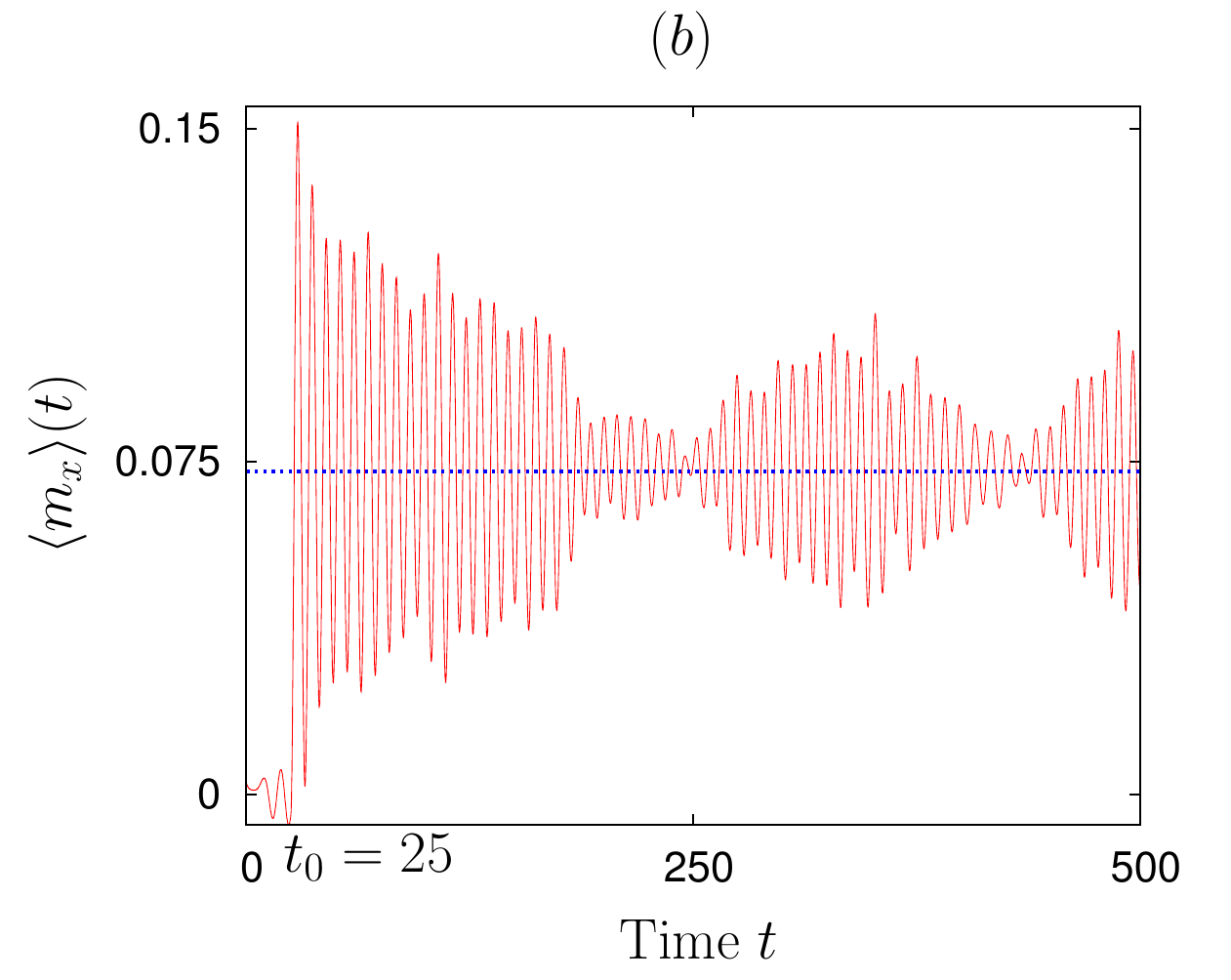}
\caption{(Color online) $\langle m_x \rangle (t)$ vs. time $t$ for the
(a) water-bag QSS, and (b) Fermi-Dirac QSS, in the
HMF model under the
action of the 
perturbation, Eqs. (\ref{perturbation-HMF}) and
(\ref{HMF-perturbation-Kt}), with $h=0.1$ switched on at time $t_0=25$. (a) The full line in the main plot
shows the result of $N$-particle simulation, while the dashed horizontal line is the theoretical time-averaged
value of $\langle m_x \rangle (t)$ given in Eq.~(\ref{timeaverage}). The system size is $N=10^5$,
while the parameter $p_0$, corresponding to energy $e=0.7$, is
approximately $1.095$.
In the inset, the numerical result (full line) 
is compared with the theoretical prediction (\ref{mx_WB_vs_time})
(dashed line). (b) The full line represents simulation results, while the horizontal
dashed line is the theoretical asymptotic value given in Eq. (\ref{mx-final-FD}). The system size is $N=10^5$,
while $\beta=10$ and $\mu=1.2$, giving energy $e \approx 0.7$.
}
\label{fig1}
\end{figure}

The oscillatory behavior of $\langle m_x \rangle (t)$ predicted for the water-bag
state, see formula (\ref{mx_WB_vs_time}), is checked in Fig.~\ref{fig1}(a). Oscillations 
around a well-defined average persist indefinitely with no damping, as predicted 
by the theory. In the inset of the same panel, the theoretical prediction 
is compared with the numerical result for a few oscillations. While the agreement is quite
good for the first two periods of the oscillations, the numerical data
display a small frequency shift with respect to the theoretical
prediction. Moreover, an amplitude modulation may also
be observed. We have checked in our $N$-particle simulations that
different initial realizations 
produce different frequency shifts, which has a consequence when averaging over
an ensemble of initial realizations, as discussed below.

In Fig.~\ref{fig1}(b), we show $\langle m_x \rangle (t)$ for the
Fermi-Dirac QSS. In this case, we have the theoretical prediction only for
the asymptotic value $\overline{m}_x$ given in Eq.~(\ref{mx-final-FD}). The time evolution 
of $\langle m_x \rangle (t)$ displays beatings and revivals of oscillations around this theoretical 
value, shown by the dashed horizontal line in the figure.
There is no sign of asymptotic convergence, even running for longer times. 
For this high value of $\beta$, which makes the Fermi-Dirac distribution very close to the
water-bag one, we cannot conclude that there will be damping in time. We have observed a
damping for smaller values of $\beta$ when the Fermi-Dirac distribution comes closer to
a Gaussian.

\subsection{Linear response of the homogeneous equilibrium state: Single
realization}
\begin{figure}[h!]
\centering
\includegraphics[width=90mm]{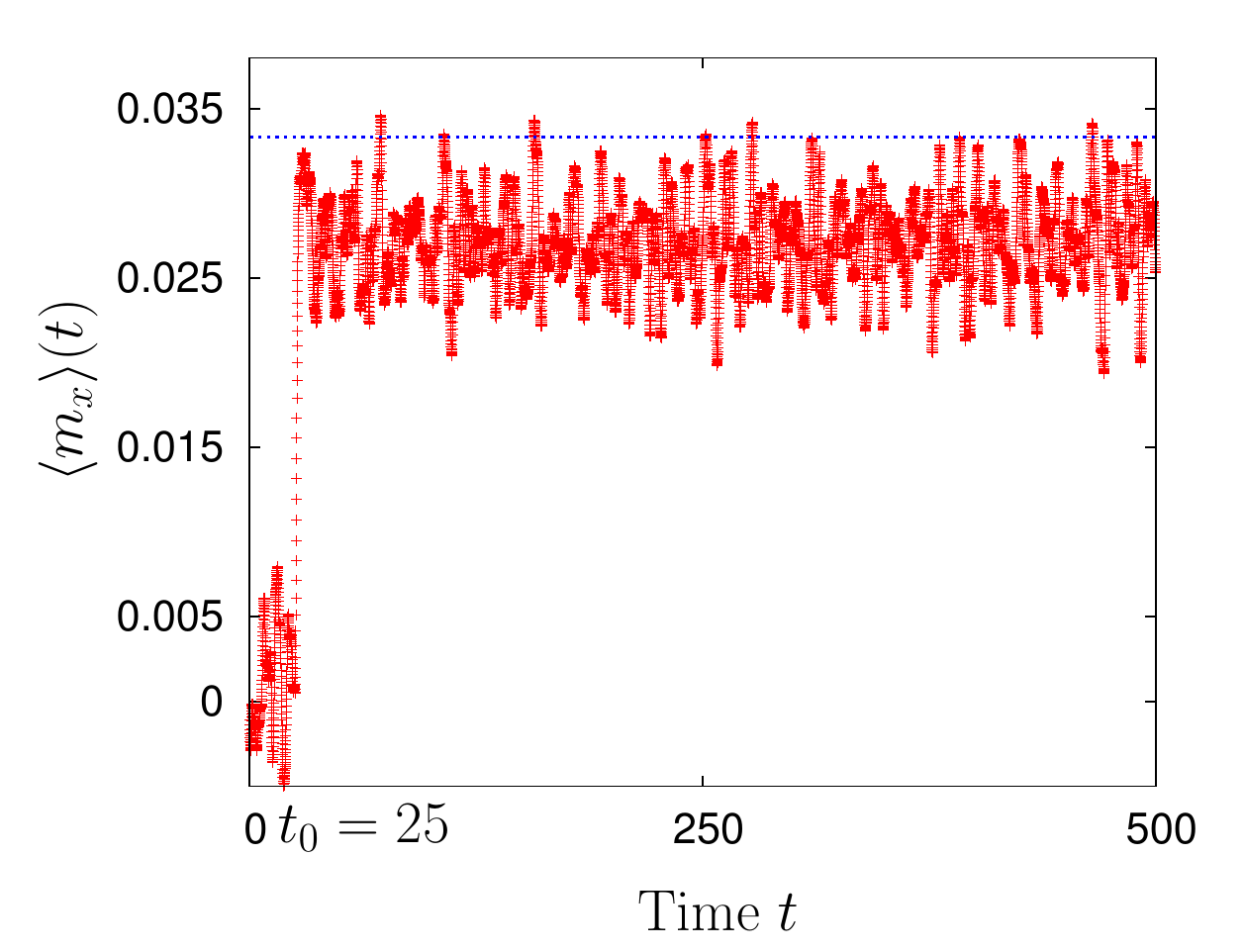}
\caption{(Color online) $\langle m_x \rangle (t)$ vs. time $t$ for the
Gaussian QSS in the
HMF model under the action of the 
perturbation, Eqs. (\ref{perturbation-HMF}) and
(\ref{HMF-perturbation-Kt}), with $h=0.1$ switched on at time $t_0=25$.
The line made of pluses represents the result of
$N$-particle simulation, while the dashed horizontal line is the theoretical asymptotic value given in 
Eq.~(\ref{mx-final-equilibrium}). The system size is $N=10^5$,
while $\beta=0.5$, so that the energy $e=1.5$.
}
\label{fig2}
\end{figure}

\l{Linear-response-equilibrium-numerical}
In Fig.~\ref{fig2}, we show $\langle m_x \rangle (t)$ for
the Gaussian QSS. After the application of the
external field, the magnetization sharply increases and fluctuates around a value which is slightly
below the theoretical prediction, Eq.~(\ref{mx-final-equilibrium}). Convergence to this latter value 
is observed on longer times.

\begin{figure*}[here!]
\centering
\begin{tabular}{lr}
\parbox[l]{9cm}{
\includegraphics[width=90mm]{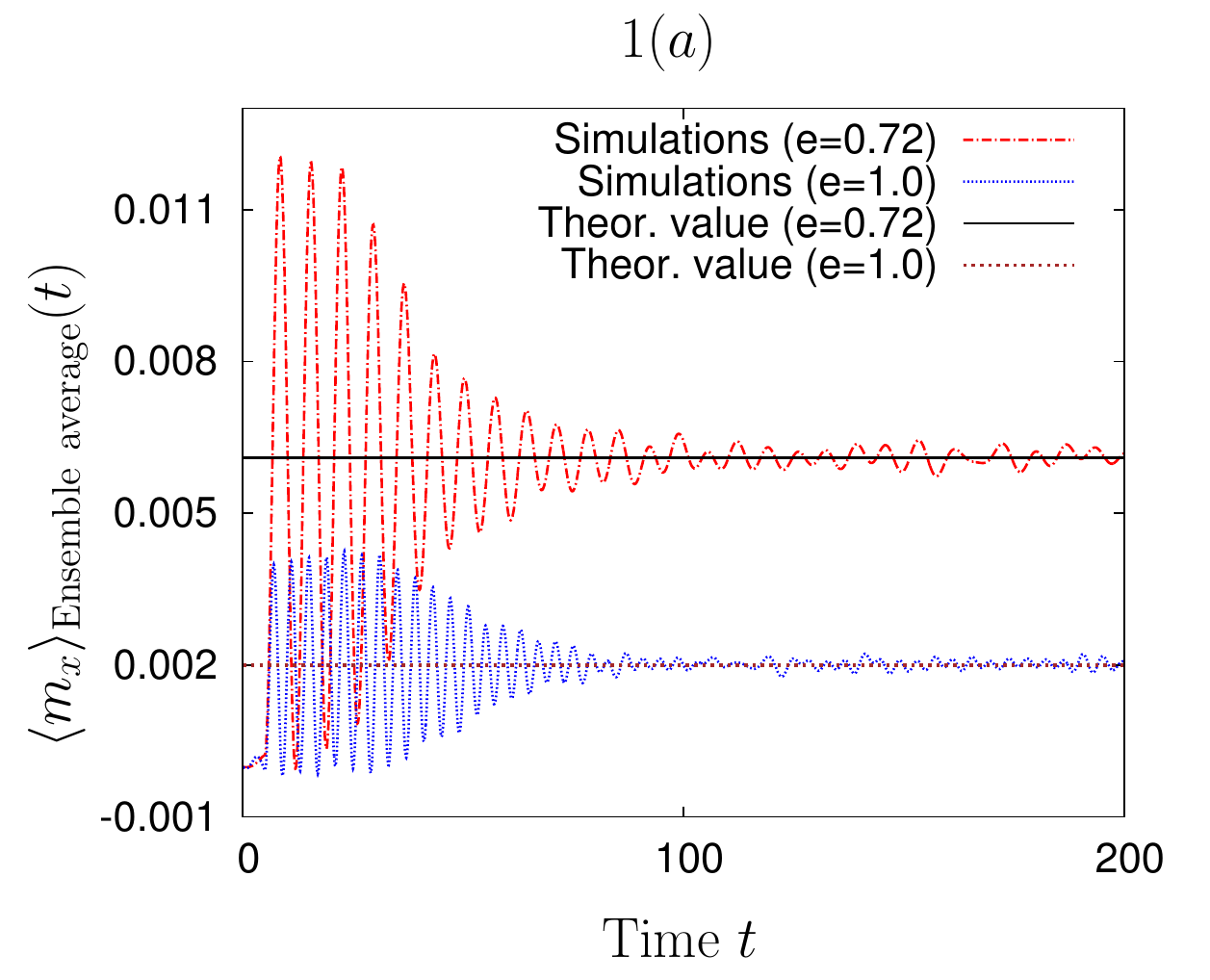}
}&
\parbox[r]{9cm}{
\includegraphics[width=90mm]{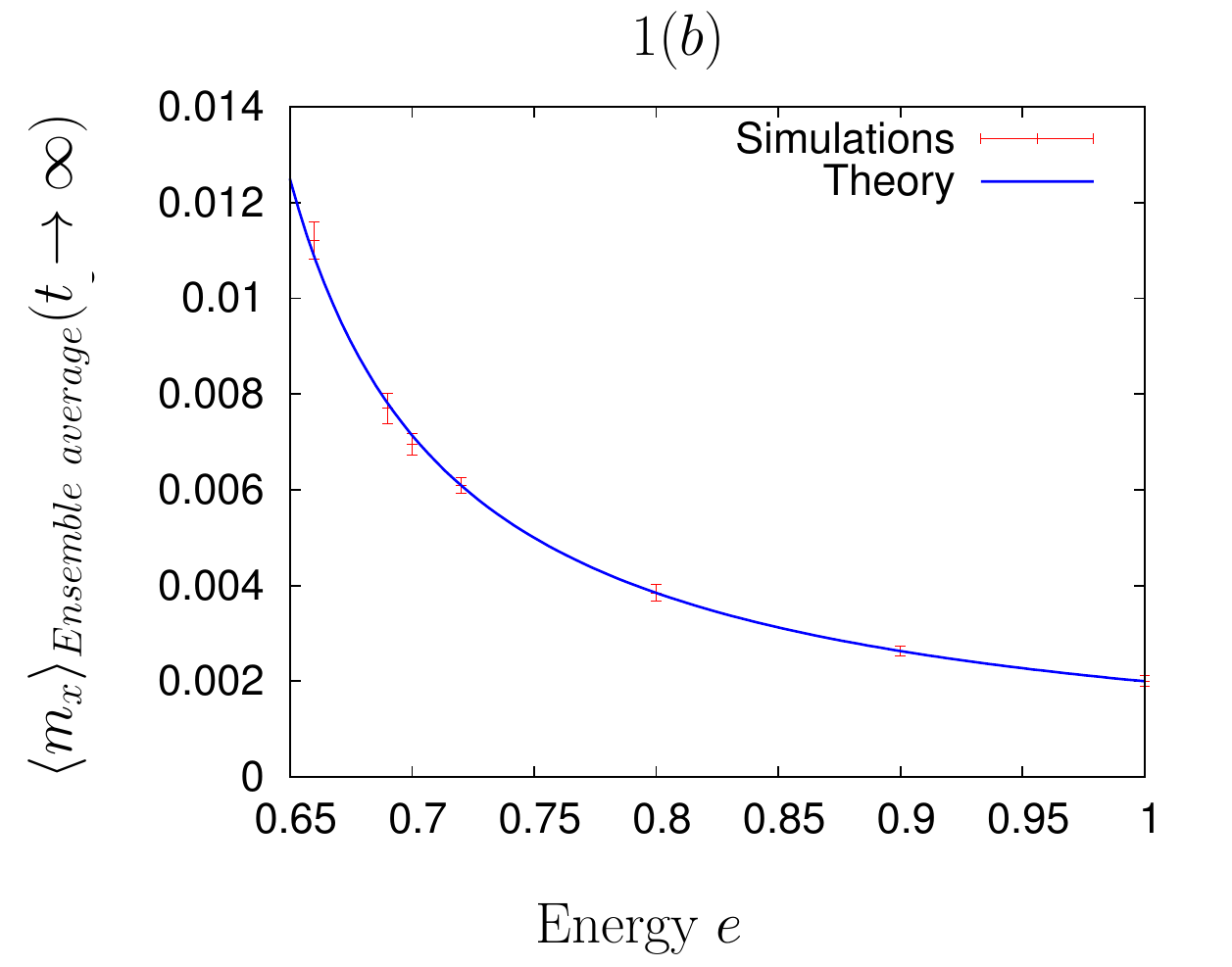}
}
\end{tabular}
\begin{tabular}{lr}
\parbox[l]{9cm}{
\includegraphics[width=90mm]{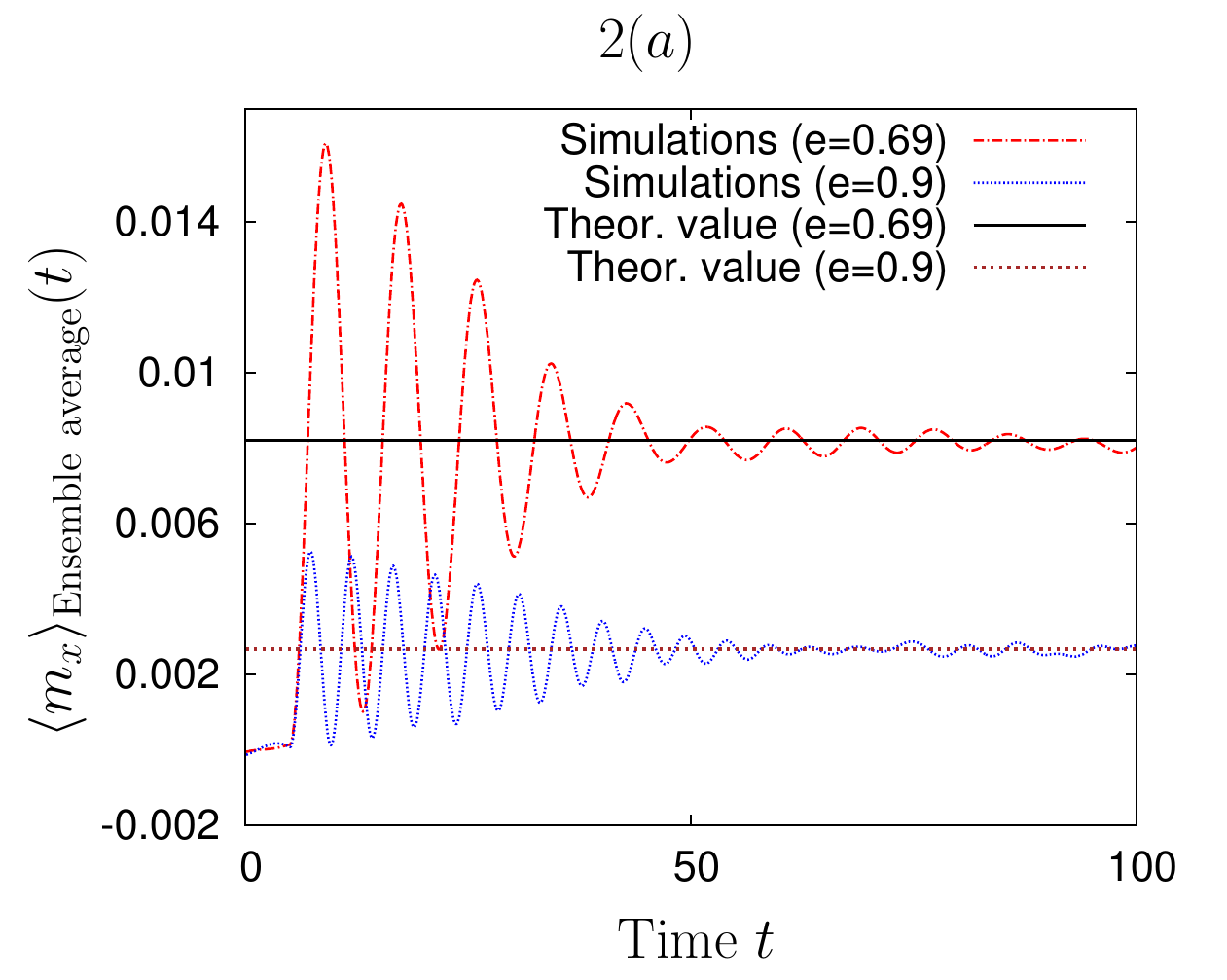}
}&
\parbox[r]{9cm}{
\includegraphics[width=90mm]{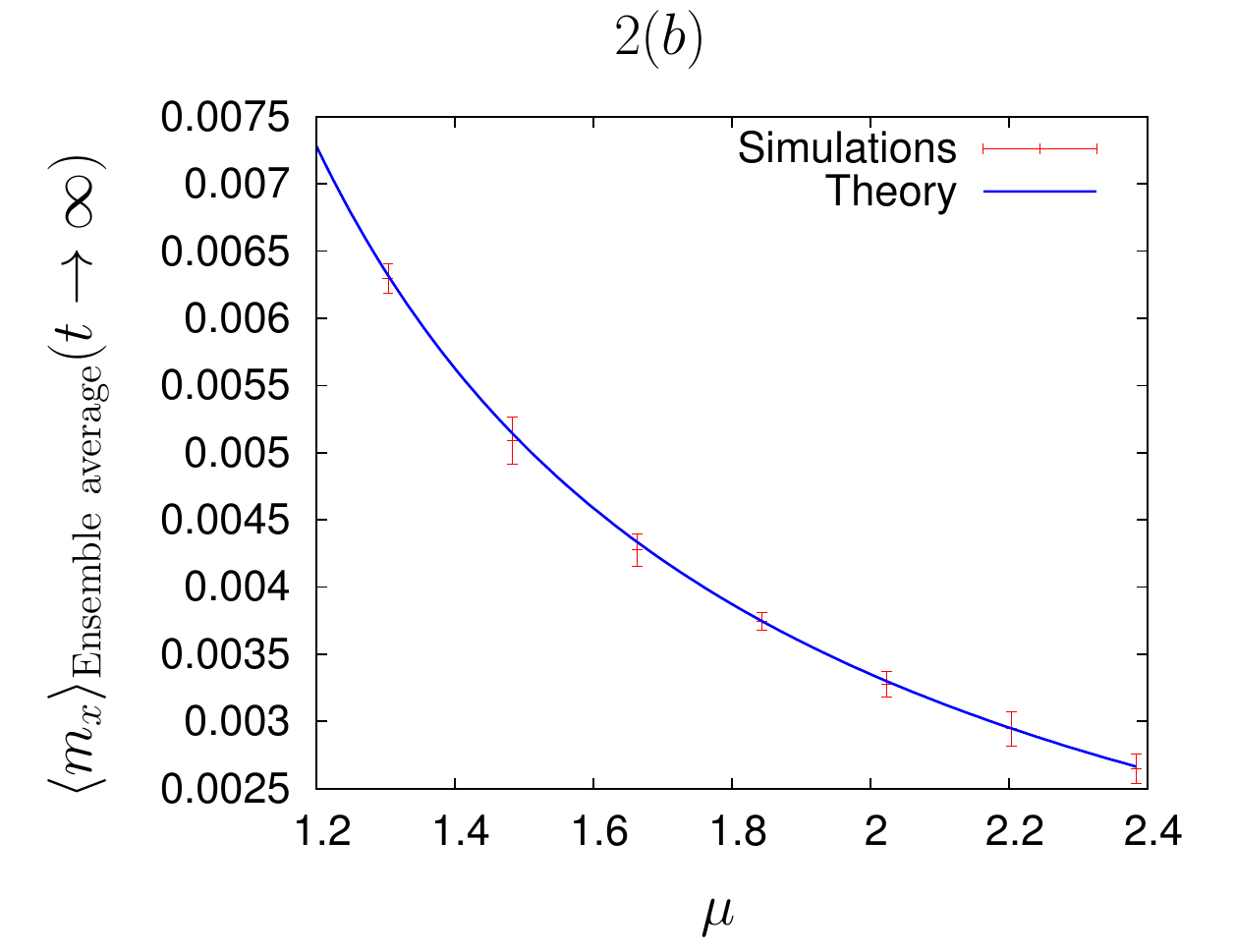}
}
\end{tabular}
\begin{tabular}{lr}
\parbox[l]{9cm}{
\includegraphics[width=90mm]{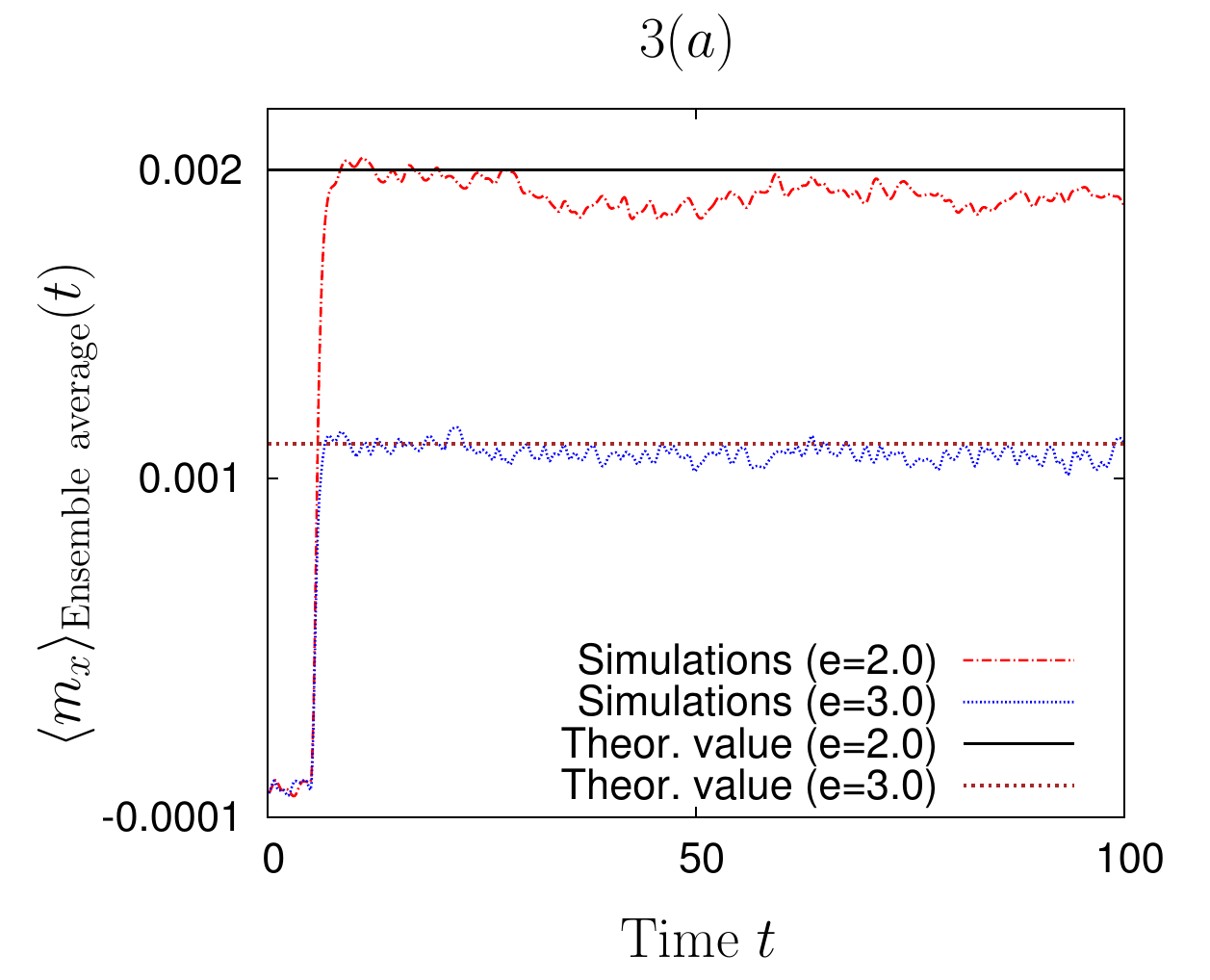}
}&
\parbox[r]{9cm}{
\includegraphics[width=90mm]{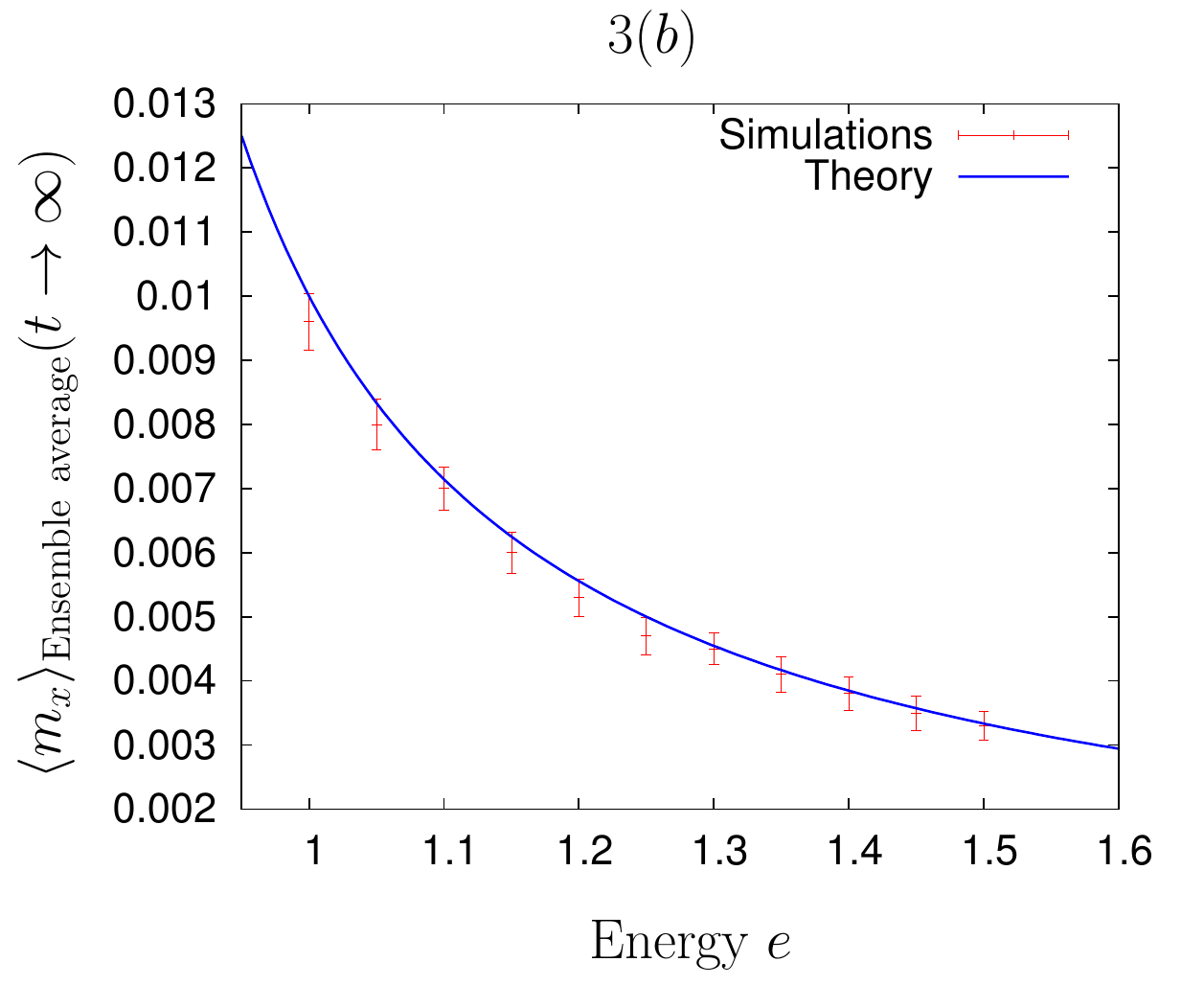}
}
\end{tabular}
\caption{(Color online) Linear response of a water-bag QSS (panels 1(a),
1(b)), a
Fermi-Dirac QSS with $\beta=10$ (panels 2(a), 2(b)), and the homogeneous equilibrium state (panels 3(a),
3(b)) for the HMF model under the
perturbation, Eqs. (\ref{perturbation-HMF}) and
(\ref{HMF-perturbation-Kt}), with $h=0.01$. 
All simulation data have been averaged over several thousand
realizations of the initial state; for details, see text. In each case, panel (a) shows the time evolution of 
the averaged magnetization $\langle m_x \rangle_{\mbox{Ensemble average}}(t)$ as
obtained from $N$-particle simulations, and its asymptotic approach either to the time average in Eq.~(\ref{timeaverage})
for the water-bag initial state or to $\overline{m}_x$ given
in Eq.~(\ref{mx-final-FD}) for the
Fermi-Dirac QSS, or to $\overline{m}_x$ given
in Eq.~(\ref{mx-final-equilibrium}) for the Gaussian QSS.
In panel (b), we show the $N$-particle simulation results for the asymptotic
magnetization  as a function of energy (the parameter $\mu$ in the Fermi-Dirac case). The error 
bars denote the standard deviation of fluctuations around the asymptotic value. The results compare well with
the theoretical predictions. The system size $N$ is $16,000$ for panels
1(a), 1(b) and panels 2(a), 2(b), and $10,000$ for panels 3(a), 3(b).
}
\label{fig3}
\end{figure*}
\subsection{Average over initial realizations}
\l{Averages}

In this section, we present numerical results for the three initial
QSSs (water-bag, Fermi-Dirac, Gaussian), obtained after
averaging the time evolution of $\langle m_x \rangle(t)$ over a set of
realizations (typically a thousand) of the initial state. We define the average
\be
\langle m_x \rangle_{\mbox{Ensemble average}} (t)= \fr{1}{N_s} \sum_{n=1}^{N_s} \langle m_x \rangle_n (t),
\ee
where $\langle \cdot \rangle_n$ labels the sample and $N_s$ is the total number of different realizations.

In all cases, we observe a relaxation to an asymptotic value. 
For the water-bag distribution, this value compares quite well with the
time-averaged magnetization given in formula (\ref{timeaverage}), see
Fig.~\ref{fig3} panels (a) and (b).  
The mechanism by which the relaxation to the asymptotic value occurs in the water-bag case,
in the absence of a true relaxation of a single initial realization, is the frequency
shift present in the different initial realizations. This leads at a given time
to an incoherent 
superposition of the oscillations of the magnetization.
For other distributions, the numerically determined asymptotic value 
is compared with the theoretical value for the single realization
$\overline{m}_x$, given in Eq. (\ref{mx-final-FD}) and Fig.~\ref{fig3} panels 2(a) and 2(b), and
formula (\ref{mx-final-equilibrium}) and Fig.~\ref{fig3} panels 3(a) and 3(b). The agreement is quite good.

\subsection{Relaxation of QSS to equilibrium}
\l{relaxation}
\begin{figure}[h!]
\begin{center}
\includegraphics[width=90mm]{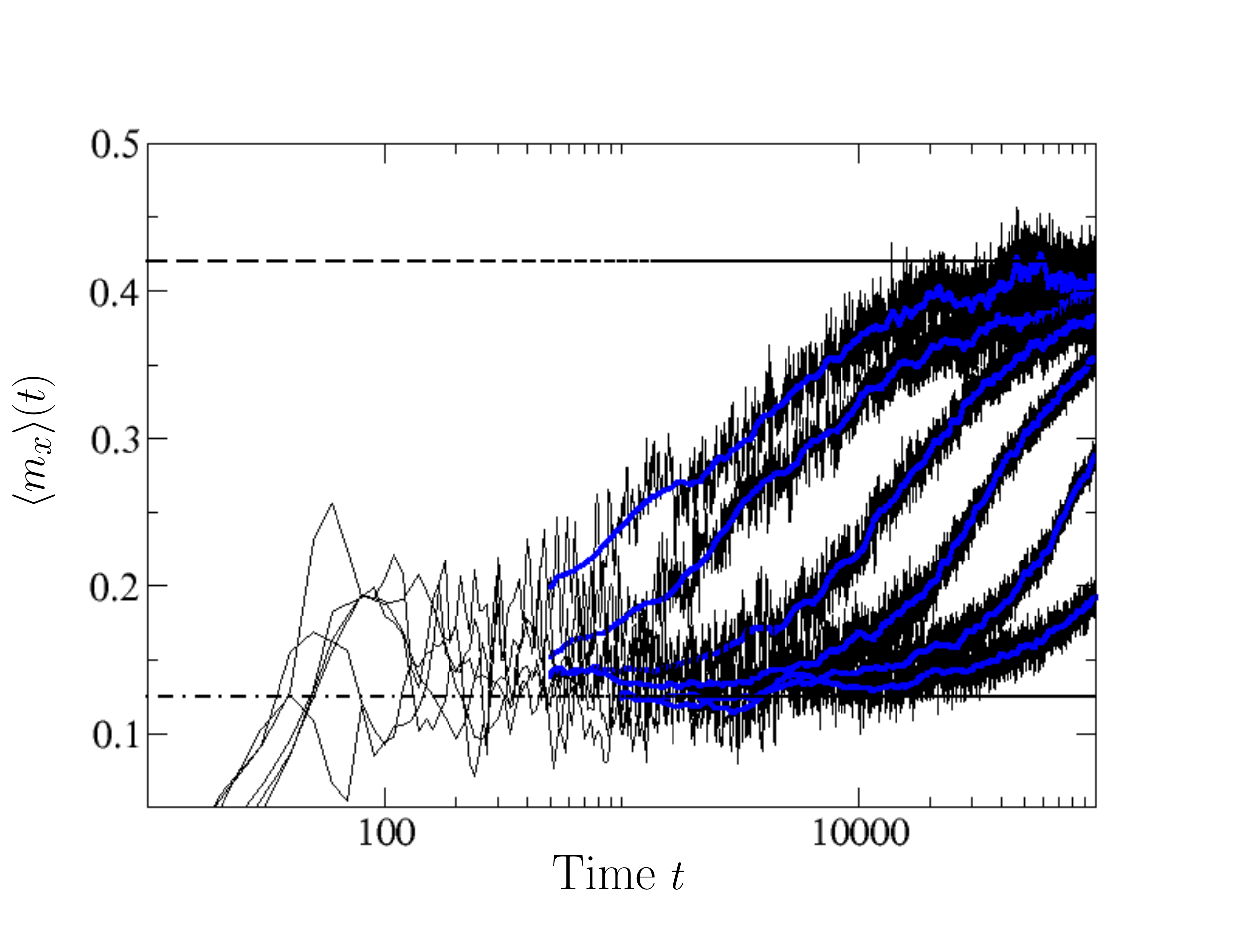}
\end{center}
\caption{(Color online) Two-step relaxation of the water-bag QSS toward the Boltzmann-Gibbs equilibrium: $\langle m_x \rangle(t)$ vs. time
$t$ for increasing system size from $N=2000$ to $N=64000$ (left to right).
Under the perturbation, Eqs.~(\ref{perturbation-HMF}) and
(\ref{HMF-perturbation-Kt}) with $h=0.01$, the water-bag initial QSS 
with $e=0.65$ relaxes to an intermediate inhomogeneous QSS with $\langle
m_x \rangle \approx 0.125$ (lower horizontal
dash-dotted line) and then to the equilibrium state with $\langle m_x \rangle
\approx 0.42$ (upper horizontal dashed line). The blue thick lines refer
to running averages performed to smooth out local fluctuations.}
\l{fig4}
\end{figure}

For finite values of $N$, the perturbed HMF system finally relaxes to
the Boltzmann-Gibbs equilibrium state. The
presence of a two-step relaxation of the initial water-bag QSS with
energy $e=0.65$, first to the perturbed Vlasov state and
then to equilibrium, is shown in Fig.~\ref{fig4} for increasing system
sizes for perturbation, Eqs.~(\ref{perturbation-HMF}) and
(\ref{HMF-perturbation-Kt}), with $h=0.01$.
The relaxation to the first magnetization plateau with value $\approx 0.125$ predicted by
the linear response theory takes place on a time of $O(1)$. The final relaxation to
the equilibrium value of the magnetization $\approx 0.42$ occurs on a timescale that increases with
system size, presumably with a power law that remains to be investigated
further.

\section{Concluding remarks}
\l{Conclusions}
In this paper, we studied the response of a Hamiltonian long-range system in a quasistationary
state (QSS) to an external perturbation. The perturbation couples to the canonical coordinates of the
individual constituents. We pursued our study by analyzing the Vlasov equation for the time
evolution of the single-particle phase space distribution. The QSSs
represent stable stationary states of the Vlasov equation in the absence
of the external perturbation. We linearized
the perturbed Vlasov equation about the QSS for weak enough external perturbation to
obtain a formal expression for the response observed in a
single-particle dynamical quantity. For a QSS that is homogeneous in the coordinate, we derived a closed
form expression for the response function. We applied this formalism to a
paradigmatic model, the Hamiltonian mean-field model, and compared the
theoretical prediction for three representative QSSs (the water-bag
QSS, the Fermi-Dirac QSS and the Gaussian QSS) with $N$-particle simulations for
large $N$. We also showed the long-time relaxation of the water-bag QSS to the Boltzmann-Gibbs
equilibrium state.


\section{Acknowledgments}
S. G. and S. R. acknowledge support of the contract LORIS
(ANR-10-CEXC-010-01). C. N. acknowledges support from the Minist\`ere
des Affaires \'Etrang\`eres. Numerical simulations were done at the PSMN
platform, ENS-Lyon. We acknowledge useful discussions with P. de Buyl,
T. Dauxois, K. Gawedski, and especially, with F. Bouchet. We thank Yoshi
Yamaguchi for discussions, for a careful reading of the manuscript, and
for letting us know of a similar project he is pursuing on linear
response.
\appendix
\section{Normalization, energy density, and stability criterion for the
Fermi-Dirac distribution [Eq. (\ref{distribution-FD})]}
\l{FD-appendix}
\underline{Normalization:}
Consider the distribution Eq. (\ref{distribution-FD}). The normalization $A$ satisfies
\be
A\int_{-\infty}^\infty \fr{dp}{1+e^{\beta(p^2-\mu)}}=1.
\ee
Changing variables and doing an integration by parts, we get
\be
2\beta A\int_0^\infty dx
~\fr{\sqrt{x}~e^{\beta(x-\mu)}}{\Big[1+e^{\beta(x-\mu)}\Big]^2}=1.
\ee
The left hand side may be written in terms of the derivative $\partial
f_{\rm FD}(x)/\partial x$ of the Fermi-Dirac-like function $f_{\rm
FD}(x)=1/[1+e^{\beta(x-\mu)}]$. We get
\be
2A\int_0^\infty dx
~\sqrt{x}\Big(\fr{-\partial f_{\rm FD}(x)}{\partial x}\Big)=1.
\l{norm-FD-1}
\ee

In the limit of large $\beta$, the derivative $\partial f_{\rm
FD}(x)/\partial x$ approaches the
Delta function: $\lim_{\beta \to \infty}\partial f_{\rm FD}(x)/\partial
x=-\delta(x-\mu)$. In this limit, we may expand $\sqrt{x}$ in a Taylor series about
$\mu$,
\be
\sqrt{x}=\sqrt{\mu}+\fr{x-\mu}{2\sqrt{\mu}}-\fr{(x-\mu)^2}{8\mu^{3/2}}+\ldots,
\ee
which on substituting in Eq.~(\ref{norm-FD-1}) gives
\be
A\Big(2\sqrt{\mu}
I_0+\fr{1}{\beta\sqrt{\mu}}\mathcal{I}_1-\fr{1}{4\beta^2\mu^{3/2}}\mathcal{I}_2+\ldots\Big)=1.
\l{norm-FD-2}
\ee
Here,
\bea
\mathcal{I}_0&=&\int_0^\infty dx~ \Big(\fr{-\partial
f_{\rm FD}(x)}{\partial x}\Big) \nonumber \\
&=&\int_{-\beta \mu}^\infty dy~\fr{e^y}{(1+e^y)^2}
\nonumber \\
&\stackrel{\beta \to \infty}
     {\longrightarrow}&\int_{-\infty}^\infty dy~\fr{e^y}{(1+e^y)^2}=1,
\eea
\bea
\mathcal{I}_1&=&\beta\int_0^\infty dx~ (x-\mu)\Big(\fr{-\partial f_{\rm
FD}(x)}{\partial x}\Big) \nonumber \\
&=&\int_{-\beta \mu}^\infty dy~\fr{ye^y}{(1+e^y)^2}
\nonumber \\
&\stackrel{\beta \to
\infty}{\longrightarrow}&\int_{-\infty}^\infty dy~
\fr{ye^y}{(1+e^y)^2} =0,
\eea
\bea
\mathcal{I}_2&=&\beta^2\int_0^\infty dx~ (x-\mu)^2\Big(\fr{-\partial f_{\rm
FD}(x)}{\partial x}\Big) \nonumber \\
&=&\int_{-\beta \mu}^\infty dy~\fr{y^2e^2}{(1+e^y)^2}
\nonumber \\
&\stackrel{\beta \to
\infty}{\longrightarrow}&\int_{-\infty}^\infty dy~
\fr{y^2e^y}{(1+e^y)^2} =\fr{\pi^2}{3}. 
\eea

Thus, to order $1/\beta^2$, we find from Eq.~(\ref{norm-FD-2}) that
\be
A\Big(2\sqrt{\mu}-\fr{\pi^2}{12\beta^2\mu^{3/2}}\Big)=1,
\ee
which gives
\be
A=\fr{1}{2\sqrt{\mu}}\Big(1+\fr{\pi^2}{24\beta^2\mu^2}\Big).
\l{norm-FD-final}
\ee

\underline{Average energy:}
The average energy density is obtained from Eq.~(\ref{HMF-e-definition}) as 
\be
e=A\int_{-\infty}^\infty dp~ \fr{p^2/2}{1+e^{\beta(p^2-\mu)}}+\fr{1}{2}.
\ee
Changing variables and doing an integration by parts, we get
\be
e=
\fr{A}{3}\int_0^\infty dx
~x^{3/2}\Big(\fr{-\partial f_{\rm FD}(x)}{\partial x}\Big)+\fr{1}{2}.
\l{energy-FD-1}
\ee

Expanding $x^{3/2}$ in a Taylor series about $\mu$ and substituting in
Eq.~(\ref{energy-FD-1}) give
\bea
e&=&\fr{A}{3}\Big(\mu^{3/2}\mathcal{I}_0+\fr{3}{2\beta}\sqrt{\mu}\mathcal{I}_1+\fr{3}{8\beta^2\sqrt{\mu}}\mathcal{I}_2+\ldots\Big)+\fr{1}{2}
\nonumber \\
&=&\fr{A\mu^{3/2}}{3}\Big(1+\fr{\pi^2}{8\beta^2\mu^2}\Big)+\fr{1}{2}.
\l{energy-FD-2}
\eea
Using Eq.~(\ref{norm-FD-final}), we find that to $O(1/\beta^2)$, the
energy density is
\be
e=\fr{\mu}{6}\Big(1+\fr{\pi^2}{6\beta^2\mu^2}\Big)+\fr{1}{2}.
\l{energy-FD-final}
\ee

\underline{Dielectric function:}
Using Eqs.~(\ref{distribution-FD}) and
(\ref{dielectric-function-definition}), we get
\be
\eps(1,0)=1-A\int_0^\infty dx~ \fr{1}{\sqrt{x}}\Big(\fr{-\partial
f_{\rm FD}(x)}{\partial x}\Big).
\l{stability-FD-1}
\ee

Expanding $1/\sqrt{x}$ in a Taylor series about $\mu$, and substituting in
Eq.~(\ref{stability-FD-1}) give 
\be
\eps(1,0)=1-A\Big(\fr{\mathcal{I}_0}{\sqrt{\mu}}-\fr{\mathcal{I}_1}{2\beta\mu^{3/2}}+\fr{3\mathcal{I}_2}{8\beta^2\mu^{5/2}}+\ldots\Big),
\ee
so that to $O(1/\beta^2)$, we get
\be
\eps(1,0)=1-\fr{1}{2\mu}\Big(1+\fr{\pi^2}{6\beta^2\mu^2}\Big),
\l{mustar-FD-appendix}
\ee
where we have used Eq.~(\ref{norm-FD-final}). 


\begin{thebibliography}{99}
\bibitem{FourierLaplace} In this paper, we adopt the following convention
to define spatial Fourier and temporal Laplace transforms: An
arbitrary function $F(q,p,t)$ of the canonical coordinate, momentum, and time has the spatial
Fourier transform
\be
\widetilde{F}(k,p,t)=\fr{1}{2\pi}\int dq ~F(q,p,t)e^{-ikq}, \nonumber
\ee
so that the inverse Fourier transform is given by
\be
F(q,p,t)=\int dk ~\widetilde{F}(k,p,t)e^{ikq}. \nonumber
\ee
The temporal Laplace transform of $\widetilde{F}(k,p,t)$ for $t>0$ is defined as
\be
\widehat{F}(k,p,\om)=\int_0^\infty dt ~\widetilde{F}(k,p,t)e^{i\om t},
\nonumber
\ee
so that the inverse Laplace transform is given by
\be
\widetilde{F}(k,p,t)=\fr{1}{2\pi}\int_C d\om
~\widehat{F}(k,p,\om)e^{-i\om
t}, \nonumber
\ee
where the Laplace contour $C$ is a horizontal line in the complex
$\om$-plane that passes above all singularities of $\widehat{F}(k,p,\om)$.
\bibitem{noteXYspins}The model may also be interpreted as a mean-field
$XY$ model of magnetism, though this identification is only superficial
in the sense that the dynamics (\ref{EOM-unperturbed}) of the
model is very different from that of $XY$ spins \cite{Gupta:2011}.
\end{thebibliography}
\end{document}